\documentclass[a4paper,12pt]{article}

\usepackage{amsmath}
\usepackage{graphicx} 
\usepackage[pdftex]{pict2e}
\usepackage[utf8]{inputenc} 
\usepackage[top=0.6in, bottom=1.25in, left=1in, right=1in, footskip=0.65in]{geometry} 
\usepackage{booktabs} 
\usepackage{hyperref} 
\usepackage{cleveref} 
\usepackage{fancyhdr} 
\usepackage{xcolor}
\usepackage{bm}
\usepackage[T1]{fontenc}
\usepackage[normalem]{ulem}
\usepackage{etoolbox}
\usepackage{braket}
\usepackage{amssymb}
\usepackage{wrapfig}
\usepackage{tikz}
\usetikzlibrary{arrows.meta}

\title{Azimuthal mode decomposition Particle in Cell algorithm for cylindrical plasma sources}
\author{
  M. Ripoli\thanks{Ph.D. student, mripoli@ing.uc3m.es},  \:
  E. Ahedo\thanks{Full professor, eahedo@ing.uc3m.es}
  \: and \: M. Merino\thanks{Full professor, mario.merino@uc3m.es} \\ \\ 
 \textit{Department of Aerospace Engineering,} \\
  \textit{Universidad Carlos III de Madrid, Leganés, 28911, Spain} \\[1ex]
}
\date{}  

\newcommand{\dd}{\mathrm{d}}
\newcommand{\pd}{\partial}

\newcommand{\w}{\omega}

\newcommand{\dfdg}[2]{\frac{\pd #1}{\pd #2}}

\newcommand{\PIF}{1D+1F}
\newcommand{\PIC}{FD}

\newcommand{\round}[1]{\left( #1\right)}
\renewcommand{\square}[1]{\left[ #1\right]}
\renewcommand{\braket}[1]{\left\langle #1 \right\rangle}

\newcommand{\fdist}{\mathcal F}

\newif\ifcomments
\commentsfalse
\commentstrue 

\ifcomments
\newcommand{\mmm}[1]{{\color{red}\textbf{#1}}}      
\newcommand{\ea}[1]{{\color{blue}\textbf{#1}}}     
\newcommand{\pfp}[1]{{\color{brown}\textbf{#1}}}    
\newcommand{\mrp}[1]{{\color{teal}\textbf{#1}}}    
\else
\newcommand{\mmm}[1]{}
\newcommand{\ea}[1]{}
\newcommand{\pfp}[1]{}
\newcommand{\mrp}[1]{}
\fi

\begin{document}

\maketitle

\thispagestyle{fancy} 


\begin{abstract}
An efficient Particle-in-Cell numerical approach to perform full-dimensional kinetic simulations of low temperature plasmas is presented. 
Taking advantage of the cylindrical geometry of most plasma sources, a Fourier mode decomposition of the fields is carried out in the azimuthal ($\theta$) direction up to a chosen maximum number of modes $N_m$. Macroparticles are pushed in all $D$ dimensions and weighed, for each mode $m$, onto a $(D-1)$ dimensional grid.
The computation of the electric field for each mode is independent and reduces to solving $(N_m+1)$ $(D-1)$ dimensional Poisson problems.
The approach brings spectral accuracy in the azimuthal direction, while the computational cost is comparable to that of a simulation with $(D-1)$ dimensions.
We verify this approach against a planar test case based on a Penning discharge, widely used for benchmarking and validation purposes in the low-temperature plasma community. Our approach allows us to reduce the 2D problem into a collection of coupled 1D problems and to naturally perform spectral analysis of the different azimuthal modes, recovering the contribution of each mode to radial transport, with a computational time saving of one order of magnitude with respect to state of the art 2D particle-in-cell codes.  
\end{abstract}


\maketitle
 
\section{Introduction}

Many plasma devices operate as partially magnetized \textbf{$E \times B$} discharges,
their application ranging from material processing to space propulsion \cite{boeu23}. 
Magnetrons, Penning discharges \cite{druy40,hoh63} and  Hall Thrusters \cite{kauf85, GOEB23, ahed11s,mazo16a} fall in this category, which also includes the under-development family of Electrodeless Plasma Thrusters \cite{bath17a}, which rely on a magnetic nozzle \cite{ahed10f, meri16g} as their accelerating stage.
In all these discharges, electrons are well magnetized and closely follow magnetic field lines, whereas heavier ions are essentially unaffected by magnetic force effects.
While the variety of devices is ample, they all share similar physical working principles,  
as well as a similar susceptibility to wave-driven  transport of charged species across magnetic field lines \cite{kaga20a}.

Simulating \textbf{$E \times B$} discharges accurately and efficiently is a topic of main interest for the low-temperature plasmas
community, as the triggering and non-linear saturation of plasma instabilities remains an open problem in these devices, which drive plasma transport across the confining magnetic field lines, causing plasma losses and reduction in deficiencies. Linear wave excitation and unstable growth in collisionless, partially magnetized plasmas have been frequently approached analytically in the literature, both from fluid \cite{moro72a,smol17,ramo21,ripo25a} and kinetic \cite{kral71,cart02b,ramo26a} models, via linear stability analyses. These works highlight how interspecies drift,
arising from the difference in magnetization between ions and electrons and sometimes necessary for thrust generation (as for Hall Thrusters), acts as the main driver behind the insurgence of instabilities in $E\times B$ discharges. These phenomena can be categorized under the comprehensive category of collisionless \textit{drift-gradient instabilities} whose most notable examples are the lower-hybrid drift instability \cite{kral71,davi75,huba76} and its simplified zero-electron Larmor radius limit version \cite{saka93}, also known as the modified collisionless variant of the Simon-Hoh instability \cite{simo63}.
Some theoretical works have extended their analytical effort to describe the non-linear stage of the wave evolution under the hypothesis of weak turbulence, using quasi-linear models \cite{ripo25a,sotn80,malk85} and semi-heuristic saturation criteria \cite{lafl16b} to quantify oscillation-driven transport and heating.

All these analytical studies are however limited to weakly inhomogeneous plasmas, and furthermore, they cannot resolve the highly nonlinear time evolution of these instabilities past the linear phase. 
Numerical integration of the plasma and field dynamics can yield insights where those techniques fall short. While fluid simulation approaches successfully resolve the main aspects of the instabilities \cite{hage00,poli26}, in many instances the complete characterization of the plasma-wave interaction ultimately requires a kinetic framework to resolve the distribution function of each plasma species, and retaining at least some of the 3D nature of the involved mechanisms. This can be approached with the electrostatic Particle-In-Cell (PIC) method, where charged macroparticles are moved independently under the action of the background magnetic field and the time-varying electric potential \cite{bird97,mari24a,garr01,guai25a,bell25}.

Virtually all existing PIC codes compute the fields through finite differences (FD) over the spatial dimensions on a grid and then interpolate them to the macroparticle positions. 
However, grid based methods scale onerously with the dimensionality of the problem, due to the need to keep the number of particles per cell $N_p$ above certain limit to prevent excessive statistical noise. In addition, the interaction between the grid (Eulerian part of the scheme) and the macroparticles (Lagrangian part) is the underlying cause of the numerical finite-grid instability \cite{huan16a}; in order to keep it checked, demanding constraints exist on the time step $\Delta t$ and the grid size $\Delta x$.

Given the large computational times that come with the standard PIC method in 3D, most of the existing analyses are in practice limited to 
2D. Many 2D PIC codes have been used to study cylindrical discharges, restricting their analyses only to  axial-azimuthal \cite{bell25,char19b,char20a} or radial-azimuthal \cite{janh18b,villa21} sections of the real device. 
Yet many other works explore the axisymmetric part of the plasma response focusing on the axial-radial plane of  the source, either ignoring or externally modeling any azimuthal fluctuations \cite{parr06a, pera22b, maty10, lewe22, mari24a,tacc08b,guai26a}.

Interestingly, the vast majority of $E\times B$ plasma devices possess a cylindrical geometry, a fact that can be exploited to establish a more efficient (and also, accurate) simulation scheme for the 3D problem than gridding over all three spatial dimensions.
Indeed, by Fourier-expanding in the azimuthal direction $\theta$ the fields that are traditionally defined over the 3D grid (electrostatic potential and field, electron and ion density and current), it is possible to approach the simulation of the full problem by solving separately each azimuthal mode $m$ up to a desired truncation mode number $N_m$ in a 2D grid in the axial-radial directions. The remaining directions are best left untouched, and follow a finite-differencing treatment, 
as this offers higher flexibility to deal with complex boundary conditions.
Similarly, a radial-azimuthal simulation can be rendered into a collection of 1D radial problems, one for each azimuthal mode, resulting in substantial computational savings.
While macroparticles must be pushed in full-dimensional space, their weighing in the azimuthal direction is carried out by projecting their density onto the Fourier basis in 
$\theta$. 
Their weighing in the the finite-difference directions can continue to be carried out on a grid using shape functions. 
The solution of each mode of the electrostatic potential, $\phi^m$, requires solving a separate
Poisson problem with the corresponding mode of the charge density.
The full-dimensional electrostatic field is readily constructed from the contribution of each mode to push the macroparticles in the next time step.

This modal decomposition technique could also enable the efficient study of discharge chamber non-symmetries. For instance, Hall effect thrusters are not strictly axisymmetric due to the presence of the external neutralizer cathode, which in many cases is displaced to one side of the device \cite{ahed11s,mazo16a}; in electrodeless plasma thrusters, the gas injector is seldom axisymmetric, and solenoid and permanent magnet manufacturing tolerances cause non-uniformities in the applied magnetic field, which produce observable asymmetries in the plasma fields \cite{inch23a}. Short from a full 3D simulation, any 2D simulation must ignore these aspects, and the impact they can have on the discharge response.

Crucially, the number of modes of interest $N_m$ in any of these practical problems is usually small.
Indeed, many plasma fluctuations, like rotating spokes, can be effectively described with a few azimuthal modes, say $<20$.
Conversely, for the study of high-frequency short-wavelength phenomena such as the Electron Cyclotron Drift instability \cite{ramo21,bell25,janh18b,cava13}, one could reduce the azimuthal $2\pi$ domain to just an arch, say $2 \pi / m_{min}$, effectively limiting the lowest azimuthal mode considered to $m_{min} < m$ and allowing to explore finer azimuthal structures.
Most importantly, a successful simulation must account for all modes $m<N_m$ 
that measurably contribute to induced plasma transport in the cross-field direction. 
This cross-field transport is caused by the phase-correlated fluctuations in the electron density $n_e$ and the electrostatic potential $\phi$, which yield a net $\bm E\times \bm B$ flow in the cross-field direction after averaging \cite{kaga20a}. Reliably predicting this transport channel is key to obtain accurate azimuthally-averaged density and potential profiles.
%
Interestingly, all studies based on axial-radial simulations, 
such as those  referenced above, can be regarded as the $N_m=0$ limit, as they merely investigate the axisymmetric plasma source response. Those efforts must necessarily introduce an additional transport parameter (referred to as anomalous transport coefficient) to emulate the effect of the contribution of the azimuthal modes to radial  $\bm E\times \bm B$ transport. Tuning this numerical parameter to obtain accurate results is a major challenge in those simulations.

Furthermore, the total number of macroparticles needed for an accurate simulation depends only on the axial-radial grid resolution and the number of azimuthal modes $N_m$. 
In addition, another advantage of the Fourier treatment of the azimuthal dimension is the spectral accuracy of the method, and  the elimination of finite grid instabilities and other grid-related numerical phenomena \cite{evst13} (if only in the $\theta$ direction).
Last but not least, it is worth noting that when analyzing standard PIC simulation results, one often ends up computing the Fourier modes of the solution a posteriori, due to their inherent interest; a plus of the presented approach is that it directly yields the desired modal amplitudes, as these are inherently employed in the computation.
All this means that
this approach has the potential to accelerate the computation of the full solution of the problem compared to traditional FD PIC.

Perhaps not surprisingly, this simulation approach is not fully new, as it has already been successfully used in laser-plasma physics where it has been termed the Particle-In-Fourier 
(PIF) method or, simply, spectral PIC \cite{evst13,lifs08,lehe16,jala17,mitc19,kirc20,shen24}. However, and to the best of our knowledge, its application to low temperature plasma physics has not been demonstrated.

This work explores and discusses the simulation approach based on this concept as outlined above, and its applicability to problems of interest in the field of low temperature plasma physics.
We present a mixed PIC/PIF architecture, based on our existing full PIC code \texttt{PICASO}. 
To illustrate the capabilities of the approach, we choose a collisionless cylindrical Penning discharge, and reduce this 2D radial-azimuthal problem into a collection of 1D radial ones.  
To this end, we employ a spectral Fourier mode discretization in the azimuthal direction and regular finite differencing (FD) in the radial direction (\PIF ). 
The chosen problem is ideally suited to demonstrate the discretization process and analyze the advantages of the algorithm, and also serves as a first step toward the treatment of 3D problems using a grid in $(z,r)$ and modal decomposition in $\theta$ (2D+1F).
An additional advantage of this choice is that robust numerical studies exist in the literature on the convergence and benchmarking of the simulation of a collisionless Penning discharge, which allows us to compare the new algorithm directly against the traditional PIC approach.
Preliminary versions of this work were presented in \cite{ripo25b} and \cite{ripo26a}.

The rest of the paper is structured as follows:
In Section \ref{sec:model} the model and implementation of the \PIF\ scheme is detailed.
The numerical constraints that apply to this simulation scheme are explained. This section also introduces the numerical configuration of the collisionless Penning discharge test case, which is simulated in Section \ref{sec:results} to verify the algorithm and its implementation. To this end, we compare against a recent benchmarking effort of the low-temperature plasma community that utilized multiple independent electrostatic PIC codes \cite{powi26}.
In section \ref{sec:convergence}, a sensitivity analysis on the effect of numerical parameters is performed, including a discussion of the simulation time.
Lastly, Section \ref{sec:conclu} presents the conclusions of this work.
%

\section{Model and numerical implementation}
\label{sec:model}
 
The model is intended to describe low temperature plasma sources using cylindrical coordinates $(r,\theta,z)$. In the present work, we limit our description to 2D problems in the $(r,\theta)$ plane, with the extension to 3D problems being straightforward. Whenever cartesian coordinates in this radial-azimuthal plane are needed, they will be referred to as $x,y$.

The distribution function of each species $\varsigma$ is discretized using markers, or macroparticles, of weights $w_p$, positions $\bm x_p = [r_p,\theta_p]$, and velocities $\bm v_p=[v_{xp},v_{yp}]$,
\begin{align}
\fdist_\varsigma\left(\bm x, \bm v, t\right) = \sum_{p \in \varsigma} w_p \delta\left(\bm x - \bm x_p(t)\right) \delta\left(\bm v- \bm v_p(t)\right) .
\label{eq:Dirac}
\end{align}
The state of each macroparticle is stored by the triplet ($r_p$, $\cos\theta_p$, $\sin \theta_p$), and its velocity in cartesian coordinates ($v_{xp}$, $v_{yp}$). This representation efficiently stores the necessary trigonometric functions of the azimuthal coordinate, needed in the algorithm later on.
As in other time-explicit PIC codes, macroparticle positions $\bm x_p$ are stored at integer time steps, and macroparticle velocities $\bm v_p$ at half-integer time steps.

Knowing the  electric field at the location of the macroparticles at time $t_{n}$, $\bm E_p^{n}$ (and any applied magnetic field $\bm B_p$, here assumed stationary),
and the macroparticle velocities at time $t_{n-1/2}$, $\bm v_p^{n-1/2}$, 
an explicit Boris leapfrog push scheme as the one outlined in \cite{delz13} is used to perform the time update. 
A cartesian coordinate push to update the positions to $t_n$ (rather than a cylindrical coordinate push) is used for numerical stability of the trajectories near the axis. 
First, the cartesian coordinates of the position are computed as $x_p=r_p\cos\theta_p$ and $y_p=r_p\sin\theta_p$; then,
\begin{align}
    \bm x_p^{n} & = \bm x_p^{n-1} + \bm v_p^{n-1/2} \Delta t.
\end{align}
Next, the velocities are updated to time $t_{n+1/2}$, relying on the definition of several intermediate auxiliary variables:
\begin{align}
    \bm v^-_p & = \bm v_p^{n-1/2} + f \bm E_p^{n},
    \\
    \bm v^+_p & = \bm v^-_p + f \bm v^- \times \bm B_p ,
    \\
    \bm v_p^{n+1/2} & = \bm v^-_p + \frac{2 f}{1 + f^2 |\bm B_p^n|^2} \bm v^+_p \times \bm B_p  + f \bm E_p^{n},
\end{align}
with $f = {q_p \Delta t}/{2m_p}$.
This time marching procedure yields the coordinates $(x_p,y_p)$  of the particle at time $t_{n}$, and the velocities $(v_{xp},v_{yp})$ at time $t_{n+1/2}$.
As such, this part of the algorithm does not differ substantially from a standard 2D FD PIC scheme.
The push is followed by the computation of $r_p=\sqrt{x_p^2+y_p^2}$, and then $\cos\theta_p$
and $\sin\theta_p$ directly from $x_p/r_p$, $y_p/r_p$.

The main differences with respect to the traditional PIC scheme arise in the particle weighting.
First and foremost, any function of $(r,\theta)$ such as the electrostatic potential $\phi$ or the electron and ion densities, $n_e$ and $n_i$, are Fourier-expanded in the $\theta$ direction of the cylindrical domain. 
For example, the electron density is decomposed into its azimuthal modes, up to a truncation mode $N_m$, as
\begin{align}
n_e(r,\theta,t) = n_e^0(r,t) + \Re\left[\sum_{m=1}^{N_m} n_e^m(r,t) \exp(i m \theta)\right]
\label{eq:fourierexpansion}
\end{align}
with $n_e^m$ the complex amplitude of mode $m$; note that this decomposition takes into account that the complex amplitude associated to positive and negative $m$ modes are complex conjugate of each other to reduce the sum to  non-negative modes only, which implicitly introduces a factor 2 in the amplitude of modes $m>0$.
This decomposition isolates the axisymmetric part of the electron density, $n_e^0$, from its non-symmetric part ($n_e^m$, $m>0$), and conveniently allows the description of rotating modes. For example, a simple $m=1$ rotating spoke is characterized by a constant real magnitude $|n_e^1|$ and a complex phase $\angle n_e^1$ that evolves linearly in time.
Note that, in reality, rotating structures can be more intricate, e.g. more than one mode may be needed to represent them, the real magnitude $|n_e^1|$ may vary in time or the phase $\angle n_e^1$ display more sophisticated dynamics, and modes may be nonlinearly coupled together. 
Second and similarly to standard PIC codes, to numerically represent the complex amplitudes $n_e^m(r)$, $n_i^m(r)$, $\phi^m(r)$, $\bm E^m(r)$ and any other function of $r$,
an equispaced grid is introduced to discretize them radial $r$ direction with grid size $\Delta r$ (but not the azimuthal one). 
In the present implementation, all quantities are computed at the grid nodes $r_i$ irrespective of their physical nature; possible extensions of this work include the use of staggered grids for the potential and the fields.

Macroparticle weighing to compute $n_e^m(r)$ and $n_i^m(r)$ for each $m$ must account for the different treatment of the azimuthal and radial  directions.
In the azimuthal direction, the contribution of each macroparticle to
the corresponding density map is computed by projecting its position and weight onto the $m$-th Fourier mode. A Dirac delta shape function is employed in this projection; other, smoother shape function choices are possible and could be used to introduce filtering at higher azimuthal modes, but they are not explored in the present work.
The evaluation of each macroparticle contribution to mode $m$ requires evaluating the complex phasors $\exp (-im\theta_p) / 2\pi$, each representing the $m$-th component of the Dirac delta shape function.
In the radial direction, weighing is performed at the grid nodes $r_i$ through the Cloud-in-Cell (CIC) method \cite{bird97,ring11}, using a linear shape function $S_r$ for macroparticle weighting on the nodes \cite{bird97}. Other shape functions are of course possible.
All in all, the $m$-th mode of the electron density at the radial node $r_i$ is computed as:
\begin{align} 
    {n}_e^0 (r_i) &=
\sum_{p}
    \frac{w_p}{2\pi r_i \Delta r}S_r(r_p,r_i),
    \nonumber
    \\
    {n}_e^m (r_i) &=
\sum_{p}
    \frac{w_p}{\pi r_i \Delta r}S_r(r_p,r_i)
    \exp (-im\theta_p) 
    \quad \text{ for $m>0$ },
\label{eq:weight_1}
\end{align}
%
where the factor 2 in the definition of modes $m>0$ has been taken into account.
Other variables, such as $n_i$, $\phi$, follow the same decomposition.
The storage of the cosine and sine of $\theta_p$ avoids the need of performing the expensive evaluation of the imaginary exponential $\exp(-i m \theta_p)$ for $m=1$, as $\exp(-i\theta_p)=\cos\theta_p - i \sin\theta_p$. For modes $m>1$
the cost of each evaluation is ameliorated by the use of the
elementary recursion formulas,
\begin{align}
\cos(m\theta_p) &= \cos[(m-1)\theta_p]\cos\theta_p - \sin[(m-1)\theta_p]\sin\theta_p,
\\
\sin(m\theta_p) &= \sin[(m-1)\theta_p]\cos\theta_p + \cos[(m-1)\theta_p]\sin\theta_p,
\end{align}
which enable the speedy computation of these transcendental functions. The weighting routine employs a standard parallel reduction scheme to safely accumulate particle contributions onto the grid.
   
To update the electric field $\bm E$ for the next time step, the Poisson equation needs to be solved. Given its linearity, each $m$ mode of the electrostatic potential $\phi$ can then be obtained independently and in parallel from the others as
\begin{align}
\nabla^2 \phi^m
\equiv 
\frac{1}{r}\frac{\pd}{\pd r} \left( r\frac{\pd \phi^m}{\pd r}\right) - \frac{m^2}{r^2}\phi^m
= - \frac{\rho^m}{\varepsilon_0} \equiv - \frac{e(n_i^m-n_e^m)}{\varepsilon_0}
\label{poisson}
\end{align}
%
where $\varepsilon_0$ is the electric permittivity in vacuum. No artificial modification of $\varepsilon_0$ is applied in this work. The electric field is then computed as:
\begin{align}
\bm E^0 &= -\frac{\pd \phi^0}{\pd r} \bm 1_r ,
\\
\bm E^m &= -
\left(\frac{\pd \phi^m}{\pd r} \bm 1_r + \frac{im}{r}\phi^m \bm 1_\theta\right) 
    \quad \mbox{for $m>0$},
\label{eq:fielddef}
\end{align}
and the full map of the field is reconstructed simply as in equation \eqref{eq:fourierexpansion}, i.e.,
\begin{align}
\bm E(r,\theta) &= \bm E^0(r) + \Re\left[\sum_{m=1}^{N_m} \bm E^m(r) \exp(i m \theta)\right].
\end{align}
%
The radial derivatives along $r$ appearing in equations \eqref{poisson}--\eqref{eq:fielddef} are evaluated via centered finite differences. 
In practice, the use of direct-iterative preconditioning to the sparse Laplace matrix renders the time to solve the entire system of Poisson's equations a small fraction of the total computational time. 

Finally, the fields $\bm E$ (and any applied magnetic field $\bm B$) are interpolated at the macroparticle positions to yield $\bm E_p$ (respectively, $\bm B_p$).
While expression \eqref{eq:fielddef} is continuous in $\theta$ and therefore can be directly evaluated at the macroparticle azimuthal position $\theta_p$, 
in the radial direction we interpolate using the same radial shape function $S_r$, employing the value of $\bm E$ at neighboring grid points. 
The evaluation of $\exp(im\theta)$ follows the same efficient recursive approach as above.

The present implementation of the \PIF\  scheme is based on a modification of our existing \texttt{PICASO} code.
\texttt{PICASO} is a finite-difference,  explicit momentum-conserving PIC code which has been successfully applied to the study of space thruster plumes and plasma sources in a variety of $2$D configurations, mainly in the axial-azimuthal plane \cite{bell25,bayo25a} and in axisymmetric cylindrical coordinates \cite{mari24a,mari25a,guai25b}. Recently, it has been benchmarked against similar PIC codes in a standardized test case \cite{powi26}.
The code is written in FORTRAN and is parallelized on CPU 
with OpenMP using a standard particle decomposition strategy. It makes use of the PARDISO MKL \cite{intel_pardiso} library for solving the sparse matrices arising from the electrostatic Poisson problem.
 
\subsection{Numerical constraints}
\label{sec:constraints}

Just like other PIC schemes, the time-explicit \PIF\ approach is subject to  constraints in the selection of its main numerical parameters: the time step $\Delta t$, the grid size $\Delta r$, the
number of particles per cell $N_p$ (or its proxy, the particle weight $w_p$) and the maximum number of azimuthal modes $N_m$.
The addition of an axial direction $z$ to make the model three-dimensional does not modify the essence of these constraints.

The first of them are common to all explicit PIC models, and emanate from 
the requirement to explicitly resolve the smallest time scale and spatial scale in the problem.
Assuming that the electron plasma frequency $\omega_{pe} = e\ n_e^{1/2}( m_e \varepsilon_0)^{-1/2}$ is the highest frequency in the problem,
that the Debye length $\lambda_{De}= (\varepsilon_0 T_e / n_e)^{1/2} e^{-1}$ is the smallest length scale , 
and that the thermal velocity of electrons $c_e$ is larger than any other velocity, the problem presents a well-defined upper limit on the timestep $\Delta t$ and the grid size $\Delta r$ \cite{BIRD91},
\begin{align}
    \Delta t \ll \min \left( \frac{1}{\w_{pe}}, \frac{2\pi \Delta r}{c_e} \right),
    \qquad
    \Delta r \ll \pi \lambda_{De}.
    \label{PIC:CFL}
\end{align} 
The latter factor in the parenthesis of the expression for $\Delta t$ (equation \eqref{PIC:CFL}) comes from the requirement that thermal electron macroparticles may not cross more than a fraction of a radial cell per time step.
Violating these conditions is known to seriously damage the simulation by numerical  instabilities \cite{huan16a}, which results from the aliasing of the energy of higher radial modes onto lower ones due to the finite grid.


In addition, in \PIF\ PIC, a new requirement exists on the maximum number of modes $N_m$, linked  to the numerical noise generated near the axis  in the azimuthal electric field, which couples to the particle velocity.
The signal-to-noise ratio in any PIC scheme, and crucially also for each mode $m$ in \PIF, scales as $N_p^{1/2}$,
as shown in \cite{lifs08}. The previous relation can be easily inferred considering that, for $N_p$ particles with the azimuthal coordinate $\theta_p$ randomly distributed along a circle following a uniform distribution,
$m\theta_p$ is equally uniformly-distributed. Hence, both $\exp(i\theta_p)$ and $\exp(i m \theta_p)$ have the same variance, namely $N_p$, and the noise in the charge density in each mode scales roughly as $\delta \rho^m \equiv e(n_i^m-n_e^m) = O (n_e^0/N_p^{1/2})$. 

This noise propagates to the electric potential and electric field, ultimately causing spurious particle acceleration. In cylindrical coordinates, this can be especially critical close to the axis, as $\bm E^m \sim \phi^m/r$.
The noise in the $m$-th component of the potential $\phi^m$ for $m\gg 1$ at the first node of the radial grid away from the axis, $r=\Delta r$, can be estimated from the Green function of the problem as (see the derivation in Appendix \ref{app:noise})
\begin{align}
    \delta\phi^m \sim \frac{\Delta r^2}{ \varepsilon_0} \frac{\delta \rho^m}{2 m} ,
    \label{eq:noise}
\end{align}
yielding for the electric field the ordering
\begin{align}
    \delta \bm E^m \sim \frac{\Delta r}{2 \varepsilon_0} \delta \rho^m.
\end{align}
A similar relation between the amplitude of numerical fluctuations of the electric field, density and grid size is found in \cite{toua22}.

The effect of these noise fluctuations on the macroparticles is a spurious acceleration, causing a numerical push on their velocity $\delta \bm v_p = (q_p\Delta t/m_q) \sum^{N_m} \delta  \bm E^m\exp(im\theta_p)$. Summing over the $N_m$ modes, in the worst case we derive the expected velocity update due to noise $|\delta \bm v_p|$ for electron particles as
\begin{align}
    |\delta \bm v_p| \sim \sqrt{\frac{N_m}{N_p}} \frac{\w_{pe}^2 \Delta r \Delta t}{2}.
\end{align}
%
When the noise in the velocity update, $|\delta \bm v_p|$, is comparable to the thermal velocity of the electrons,  $c_e=(T_e/m_e)^{1/2}$, the population spuriously heats up, as their random velocity at the next timestep is dominated by $\delta \bm v_p$. The jump in temperature due to noise is of the order of
\begin{align}
    \frac{\delta T_e}{T_e} \sim \frac{|\delta \bm v_p|}{c_e} .
\end{align}
The constrain $|\delta v_p| \ll c_e$ avoids this particular form of numerical heating  and allows us formulate the following limiting criterion for $N_m$:
\begin{align}
    \sqrt{N_m} \ll \frac{2 \lambda_{De}}{\Delta r} \frac{ \sqrt{N_p}}{\w_{pe} \Delta t} ,
    \label{Nm:lim}
\end{align}
which reveals the following scaling properties: a) a finer spatio-temporal grid allows for higher modes to be included in the system, and b) one must increase $N_p$ linearly with $N_m$.





 
\subsection{Collisionless Penning discharge test case}
\label{sec:Penning}
 
\begin{figure}
\centering
\begin{tikzpicture}[scale=0.8]



\begin{scope}
    \def\R{3}        
    \def\r{1.1}      

    
    \draw[black, thick] (3,3) circle (\R);

    \draw[black, dashed] (3,3) circle (\r);
    
    \draw[-{Latex}, black] (3,3) -- ({\R*cos(90)+3},{\R*sin(90)+3});
    \node at ({0.7*\R*cos(90)+3},{0.7*\R*sin(90)+4.3}) {$R$};

    \node at (0,5.5) {$\phi=0$};
    
    \draw[-{Latex}, black] (3,3) -- ({\r*cos(40)+3},{\r*sin(40)+3});
    \node at ({0.7*\r*cos(40)+4},{0.7*\r*sin(40)+0.2+3}) {$r_{{inj}}$};
    
    
    \node at (2.9,2.4) {$I_{e},\, I_{i}$};
    

\end{scope}
\end{tikzpicture}
\caption{Geometry of the Penning discharge test case considered in this work. The domain is a cylinder of radius $R$, initially empty, with a uniform perpendicular magnetic field $\bm B$ coming out of the page and a fixed potential $\phi=0$ at its boundary. Electrons and ions are injected uniformly in the disk $r<r_{inj}$. Any particles reaching the edge of the domain are removed from the simulation.}
\label{fig:BCs}
\end{figure}
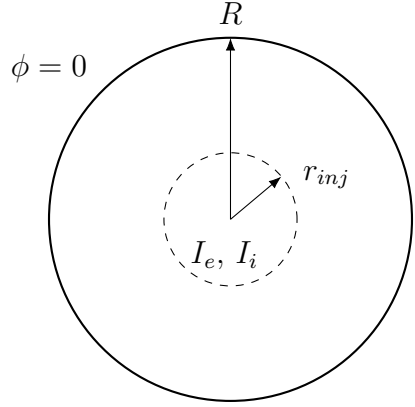

To illustrate the performance of the algorithm, we model a simple Penning discharge as an infinite cylinder of radius $r=R$, and consider a radial-azimuthal ($r,\theta$) cross-section as displayed in figure \ref{fig:BCs}. 
All variables are assumed to be uniform in the out-of-plane $z$ direction.
The discharge is modeled as collisionless, and chosen amounts of electrons and ions per unit time, $I_{e0}$ and $I_{i0}$, are injected uniformly in the central region ($r\leq r_{inj}$).   
A constant, uniform magnetic field $\bm B$ is applied in the axial direction (i.e., perpendicular to the plane of the cross section).
The radial edge of the domain is set to a constant potential $\phi=0$, and any ions and electrons reaching it are assumed to recombine and leave the system.
Initially, the domain is empty. After a fill-up transient, the simulation evolves toward a quasi-steady regime, which presents a rotating spoke.
Table \ref{tab:penning_case} lists the physical parameters of the simulation. The same  parameters as in the benchmark study presented in \cite{powi26} have been used to facilitate comparison in section \ref{sec:results}.

\begin{table}[]
    \centering
    \begin{tabular}{|c|c|c|c|}
        \hline
        \textbf{Physical parameter} & \textbf{Symbol} & \textbf{Value} & \textbf{Unit} \\
        \hline
        \hline
        Helium-$4$ ion mass & $m_i$ & $7291.2$ & $m_e$ \\
        \hline
        Domain radius & $R$ & $2.5$ & cm \\
        \hline
        Injection radius & $r_{inj}$ & $0.5$ & cm \\
        \hline
        Electron injection current & $I_{e}$ & $-20 \times 10^{-3}$ & A/m \\
        \hline
        Electron injection temperature & $T_e$ & $15$ & eV \\
        \hline
        Ion injection current & $I_{i}$ & $8 \times 10^{-3}$ & A/m \\
        \hline
        Ion injection temperature & $T_i$ & $0.025$ & eV \\
        \hline
        Applied magnetic field strength & $B$ & $0.01$ & T \\
        \hline
        Total simulation time & $t_{tot}$ & $500$ & $\mu$s \\
        \hline
    \end{tabular}
    \caption{Physical parameters of the Penning discharge test case.
    }
    \label{tab:penning_case}
\end{table}

The outer boundary of the domain $r=R$ is treated as a perfect grounded conductor, imposing on the potential the Dirichlet condition for all modes $m$ including $m=0$,
\begin{align}
    \phi^m |_{r=R}=0 . \label{BC}
\end{align}

The boundary condition at the axis, in contrast, differs for the axisymmetric mode ($m=0$) and the non-symmetric modes ($m>0$). Geometric consistency demands that $\phi^0$ have null radial derivative at the axis to avoid $E_r^0$ to be ill-defined at the centre of the circular domain, 
yielding a von Neumann condition,
\begin{align}
    \dfdg{\phi^0}{r} \bigg|_{r=0} = 0.
\end{align}
For all other modes, any dependence on $\theta$ should vanish at the axis, which imposes the   Dirichlet condition, 
\begin{align}
    \qquad \phi^m |_{r=0} = 0 \quad \text{for} \ m>0 .
\end{align}

Regarding the boundary conditions on the macroparticles,
those reaching the outer domain radius are removed from the simulation. No boundary condition on the macroparticles is needed at the axis, where the (cartesian) macroparticle kinematics and dynamics are consistently solved.

Particle injection in the core region $r<r_{inj}$ is carried out as follows.
The coordinates of newly injected macroparticles are obtained by sampling two random numbers $x_1,x_2$ uniformly distributed between $0$ and $1$, $x_1,x_2 \sim \mathcal{U}(0,1)$, so that
\begin{align}
    r_p= r_{inj} \sqrt{x_1} , \qquad \theta_p = 2 \pi x_2 .
\end{align}
This provides an axisymmetric injection in the domain. 
Macroparticles are injected following a Maxwellian distribution in velocity space with a chosen temperature for each species and zero fluid velocity.

The chosen values of the injection currents,  $I_e$  and $I_i$, the time step $\Delta t$, and the macroparticle weight $w_p$ mean that a noninteger number of macroparticles may need to be injected every time step, exceeding an integer number by a fractional reminder.
To resolve this minor circumstance and adjust the injection book-keeping precisely, each time step an additional random number is drawn and compared against the fractional remainder. If the fractional remainder is larger than the random number, an additional macroparticle is injected, otherwise the fractional remainder is accumulated to the number of macroparticles to be injected at the next step.  

\section{Results}
\label{sec:results}

A first simulation of the collisionless Penning discharge problem defined in Section \ref{sec:Penning} is presented next. The numerical parameters of this  `reference' simulation (denoted simulation R) are displayed in table \ref{tab:sim_params}, where we also define the symbols for the number of modes $N_m$, radial cell size $\Delta r$, time step $\Delta t$, and average number of macroparticles per cell $N_p$ (observe that this number varies across the domain, with a lower macroparticle count in the first cells near the axis, around $200$). 
These parameters satisfy both the grid-related constraints and the spectral ones presented in section \ref{sec:constraints}.
The simulation results are then compared against the benchmark results of \cite{powi26}.
The small cell size and time step, the large number of macroparticles per cell, and the high $N_m$ result in a high-fidelity simulation of the problem, as further verified in section \ref{sec:convergence}.


\begin{table}[]
    \centering
    \begin{tabular}{|c|c|c|c|}
        \hline 
        \textbf{Numerical parameter} & \textbf{Symbol} & \textbf{Value} & \textbf{Unit} \\ 
        \hline
        \hline
        Number of $\theta$ modes & $N_m$ & $20$ & - \\
        \hline
        \hline
        Number of cells in $r$ direction & $N_r$ & $257$ & - \\
        \hline
        Cell size & $\Delta r$ & $\simeq 0.01$ & cm \\
        \hline
        \hline
        Macroparticle weight & $w_p$ & $2 \times 10^6$ & particles \\
        \hline
        Macroparticles per cell (average) & $N_p$ & $7\times 10^3$ & macroparticles \\
        \hline
        Total macroparticles (approx) & $N_pN_r$ & $1.8 \times 10^6$ & macroparticles \\
        \hline
        \hline
        Time step size & $\Delta t$ & $0.02$ & ns \\
        \hline
        Number of time steps & $N_{t}$ & $2.5\times 10^7$ & - \\
        \hline
        \hline
        Printing time step & $\Delta t_{pr}$ & $0.05$ & $\mu$s \\
        \hline
        Number of print steps & $N_{pr}$ & $ 10^4$ & - \\
        \hline
        \hline
        Simulation wall time & $t_{wall}$ & $110$ & h \\
        \hline
    \end{tabular}
    \caption{Numerical simulation parameters in the reference case (simulation R). The simulation has been run on 16 cores of a single Intel Xeon Gold Y6542, with a RAM of $256$ GiB.}
    \label{tab:sim_params}
\end{table}

\begin{figure}
    \centering
    \includegraphics[width=0.8\linewidth]{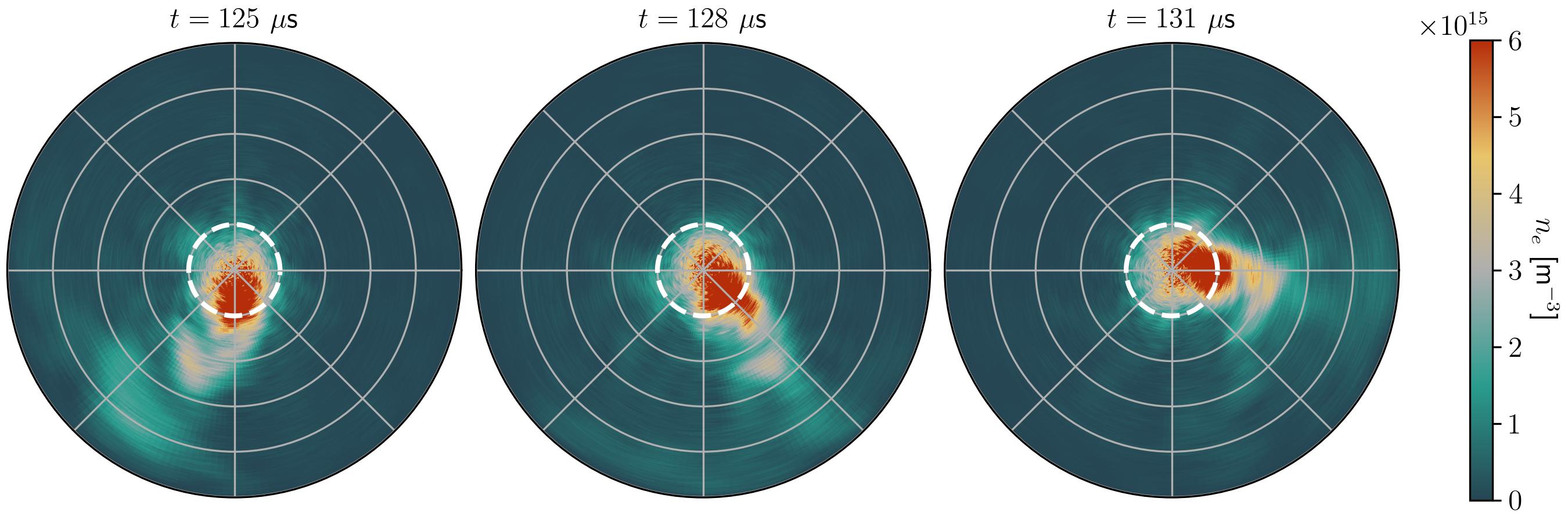}
    \caption{$2$D reconstruction of the electron density $n_e$ at three different time instants for the reference simulation case R. The white dashed line delimits the injection region.
    }
    \label{fig:rfrnc_2D}
\end{figure}

Three 2D snapshots of the electron density, after the initial transient setup is well past, are reconstructed using equation \eqref{eq:fourierexpansion} and shown in figure
\ref{fig:rfrnc_2D}. A coherent and stable anticlockwise rotating spoke structure is recovered, as expected in this parametric regime \cite{powi26}. The spoke rotates in the $\bm E\times \bm B$ direction, with
a frequency of $48$ kHz, estimated from the frequency domain analysis of section \ref{sec:freqdomain}.
The low speed dynamics mean that the electron density closely matches the ion density everywhere and, except in a thin sheath around the outer boundary, the plasma is seen to remain essentially quasineutral in the domain, again fulfilling with the expectations of this type of instability.

\subsection{Long-term averages}\label{sec:long-term}

\begin{figure}
    \centering
    \includegraphics[width=.4\linewidth]{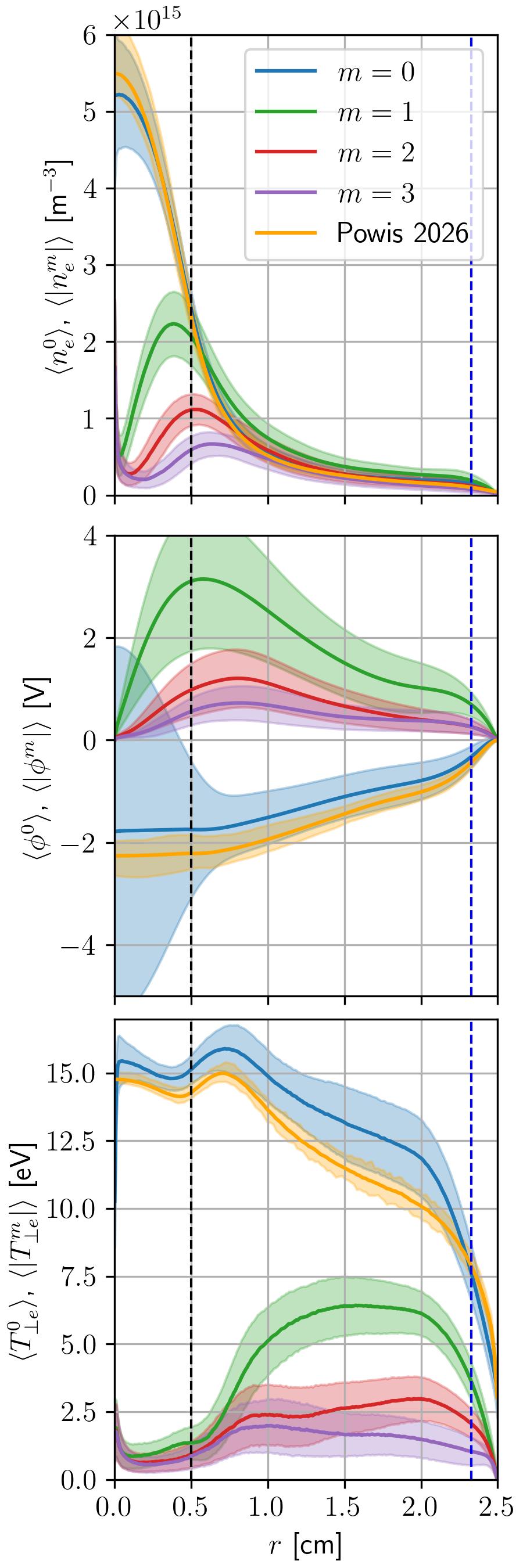}
    \includegraphics[width=0.4\linewidth]{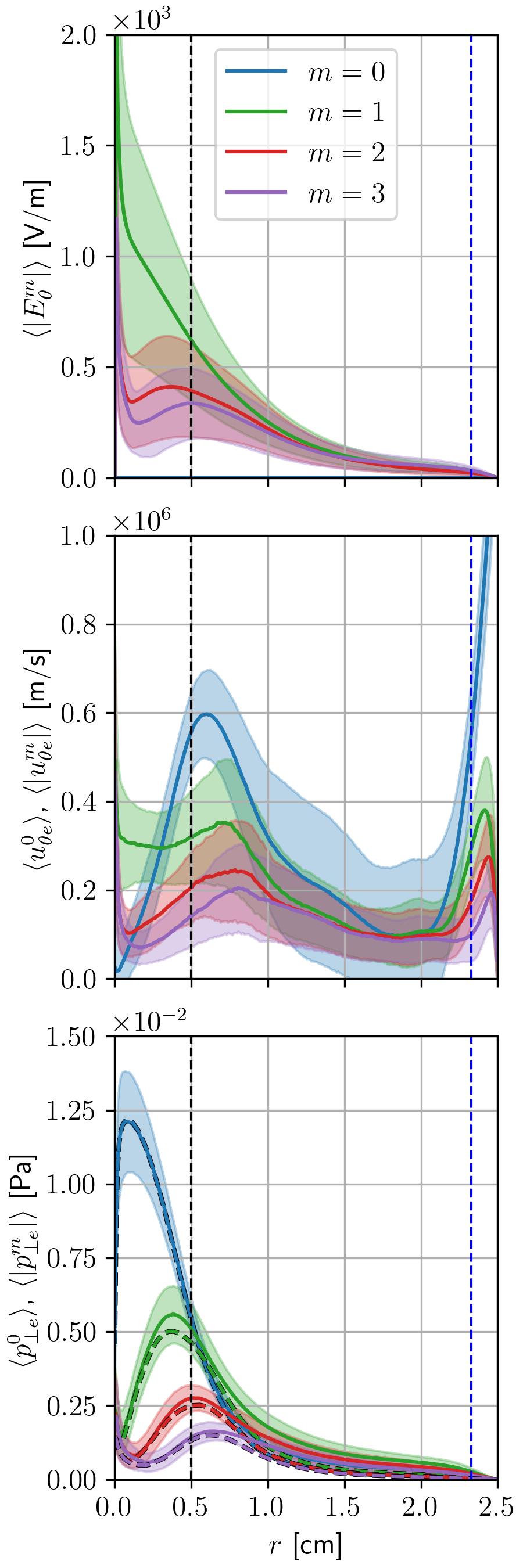}
    \caption{Long-time averages and standard deviations of the absolute value of
    $n_e^0$, $\phi^0$, $T_{\perp e}^0$, $u_{\perp e}^0$, $p_{\perp e}^0$ and $|n_e^m|$, $|\phi^m|$, $|T_{\perp e}^m|$, $|E_\theta^m|$, $|u_{\perp e}^m|$ and $|p_{\perp e}^m|$   
    for $m=1,2,3$, for the reference simulation R. The definition of long-time mean $\braket{\cdot}$ and its standard deviation have been presented in Eq. \eqref{eq:longtime}.
    Mode $m=0$ has a real amplitude (i.e. no complex phase), and in the case of $\phi$, its signed profile is shown. 
    Solid lines correspond to mean values, while the half widths of the shaded areas represent the associated standard deviation.
    The mean and dispersion of the mode zero of the benchmark simulations in \cite{powi26} are also shown for comparison.
    The dashed black vertical line denotes $r=r_{inj}$;
    blue a Larmor diameter, $r=R-2\ell_e$.
    This is computed computed with $T_e=12.5$ eV.
    }
    \label{fig:rfrnc_1D_ABS}
\end{figure}

Figure \ref{fig:rfrnc_1D_ABS} presents the long-time averages and standard deviations of the axisymmetric component of the discharge, i.e.,
$n_e^0$, $\phi^0$, $T_{\perp e}^0$, $u_{\theta e}^0$ and $p_{\perp e}^0$. 
It also depicts the absolute value of the amplitudes $|n_e^m|$, $|\phi^m|$, $|T_{\perp e}^m|$, $|E_\theta^m|$ $|u_{\theta e}^m|$ and $|p_{\perp e}^m|$ for $m=1,2,3$. 
The long-term statistics are computed
using $N=5000$ equispaced snapshots of the simulation, each separated by $2500$ timesteps from the next, at the end of the simulation, and therefore are representative of the last  $250\ \mu$s, well after the initial simulation transient (which is completed after roughly 100 $\mu$s), and covering several revolutions of the spoke.
We denote the long-term mean of each quantity with angular brackets  as $\langle\cdot\rangle$ and its standard deviation as $\sigma(\cdot)$. For instance, for $n_e^0$ this is:
\begin{align}
\langle n_e^0 \rangle = \frac{1}{N}\sum_n^N n_e^0,
\qquad
\sigma(n_e^0) = \sqrt{ \frac{1}{N-1}\sum_n^N \round{n_e^0 - \braket{n_e^0}}^2} .
\label{eq:longtime}
\end{align}
The averaging for the complex amplitudes of nonsymmetric modes ($m>0$) must be computed over their absolute values; otherwise, if the average was taken over the complex values itself, the large phase variation in this time windows would make the averages go to zero. 
The modal components of $u_{\theta e}$, $p_{r e}$ and $p_{\theta e}$, 
being themselves ratios of moments of the electron velocity distribution function,  are computed as follows:
\begin{align}
    u_{\theta e}^m(r) & = \frac{a_m}{\pi} \int_0^{2\pi} \dd\theta \exp \left( i m \theta \right) \square{\frac{g_{e,\theta}}{n_e}}(r,\theta),
    \\
    p_{re}^m (r) & = \frac{a_m}{\pi} \int_0^{2\pi} \dd\theta \exp \left( i m \theta \right)   \square{\mathcal{M}_{rre}  - m_e \frac{g_{re}^2}{n_e}} (r,\theta),
    \\
    p_{\theta e}^m (r) & = \frac{a_m}{\pi} \int_0^{2\pi} \dd\theta \exp \left( i m \theta \right)  \square{\mathcal{M}_{\theta\theta e} 
    - m_e \frac{g_{\theta e}^2}{n_e}} (r,\theta),
\end{align}
with $a_m=1$ for $m>0$ and $a_0=1/2$. 
We also define $p_{\perp e}^m(r) = [{p_{re}^m(r)+p_{\theta e}^m(r)}]/{2}$.
These definitions imply first a reconstruction in ($r$,$\theta$) of  $\bm g_{e}$  and $\mathcal{M}_{e}$, the flux vector and the complete momentum tensor of the distribution. 
This reconstruction takes into account all computed modes up to $m=N_m$.
The definitions of $T_{r e}$, $T_{\theta e}$,  and $T_{\perp e}$ are likewise.
The same quantities can be defined for ions simply through the index substitution $e\to i$.
Figure \ref{fig:rfrnc_1D_ABS} also depicts the width of the Larmor diameter, defined as $(R-2\ell_e)$ with $\ell_e=(m_e T_{\perp e}^0)^{1/2} / eB$.

These long-term averages serve as a representative depiction of the quasi-steady state of the discharge, and are the commonly computed and presented quantities in other works \cite{powi26}.
However, and as discussed later in section \ref{sec:instantaneous},
they necessarily disregard the phase relations between variables, and
they also smooth out certain physical low frequency oscillations present in each mode.
Indeed, and as can be inferred from the results in that section, the variability captured by the standard deviation is mainly indicative of the low frequency oscillations of the modes, as the noise in the simulation (which scales as $1/\sqrt{N_p}$ as discussed in section \ref{sec:constraints}), is typically smaller.

The $0$-th mode, i.e. the axisymmetric plasma response, exhibits a centered peak in density $n_e^0$ and a shallow well of the potential $\phi^0$. This is qualitatively consistent with a Penning discharge with $I_e>I_i$, where a negative radial electric field $E_r<0$ forms to push outward the excess electrons and confine the lower amount of ions, where the currents reaching the outer wall must be, in average over time, equal to the respective injection currents in steady state. The axisymmetric component of the (perpendicular) electron temperature $T_{\perp e}^0$  
is roughly uniform in and near the injection region, and gradually decays radially at a roughly constant rate until it reaches the Larmor layer at the edge of the domain, where it drops suddenly.

In the work by Powis et al. \cite{powi26}, the same Penning discharge simulation was simulated by different research groups using various 2D PIC codes, showing good agreement among them.
The left column of Figure \ref{fig:rfrnc_1D_ABS} displays, on top of our simulation results, also the mean and the variability across those benchmark simulations for the axisymmetric profiles ${ n^0_e}$, ${ n^0_i}$, ${\phi^0}$ and ${ T^0_{e}}$.
Our simulation shows an overall good match with the benchmark.
The trends and behavior of these variables are identical; differences are small and add up to a slightly lower plasma density peak at the axis, a shallower potential well, and a larger electron temperature.
%
An additional discrepancy is related to the spoke frequency,  $48$ kHz in our simulation as indicated above,
and $43.35 \pm 2.9$ kHz in the benchmark (in average over the ensemble of simulations). This discrepancy is however within the uncertainty of our own frequency estimate,
which is $\pm 8$ kHz (half the frequency bin size, as detailed in section \ref{sec:freqdomain}).

These minor discrepancies can be attributed, at least in part, to a noteworthy difference between the benchmark simulation setup and our own: while the former all use a square domain, in our work the domain is circular, as indicated above.
This is by no means a small change in the simulation setup, as in the benchmark simulations the rotating structures encounter the boundary condition at a varying radius as they go around the Penning discharge, and this can have an impact on the resulting profiles and the observed spoke frequency.

It is also worth  pointing out some other numerical differences that exist between the two works, whose effects ought to vanish in the consistency limit ($\Delta t\to 0$, $\Delta r\to 0$, $N_p\to \infty$, $N_m\to \infty$) but which may persist at finite time/space/kinetic/modal resolution.
Firstly,
we are intentionally truncating the azimuthal modal expansion in the reference simulation at $N_m$; a fixed azimuthal mode resolution allows for a better spatial resolution close to the axis, which deteriorates as $1/r$ moving out in the radial direction. In contrast, the benchmark simulations, with their 2D cartesian grid, offer the same resolution throughout the domain. It is worth noting, of course, that the benchmark simulations do not resolve an infinite number of azimuthal modes either, and moreover, finite differencing incurs in dispersion errors and aliasing for modes whose wavelength is near the grid size.
Secondly,
while 
the total number of macroparticles
    in our simulation is lower than in the 2D finite-differencing codes of the benchmark, our number macroparticles per cell, $N_p$, is actually greater. This  numerical parameter determines the statistical noise in the simulation, which suggests our simulations suffer less from this phenomenon, which can ultimately affect radial transport artificially.
Lastly,
the macroparticles and the underlying grids have different effective shapes in the two schemes. As a result, the two codes likely experience noise and numerical diffusion differently. Arguably, our cylindrical radial-azimuthal decomposition is better aligned with the principal directions of the problem than a square grid; it is conceivable that this allows a better accuracy under finite numerical parameters than a cartesian 2D grid.
 
While our approach is ill suited to reproduce square domains, and while the radial-azimuthal direction alignment is inherent to it,  the stability of our results in what pertains the numerical parameters $\Delta t$, $\Delta r$, $N_p$, $N_m$ is discussed in section \ref{sec:convergence}.
 
We next explore the behavior of the non-symmetric azimuthal modes $m>0$ in our simulation.
A key aspect of our  approach is that we truncate at a chosen mode number, $N_m$. For this truncation to be meaningful, we require that the `physical relevance' of each azimuthal mode decays with $m$. 
The nonsymmetric modes, illustrated here (in figure \ref{fig:rfrnc_1D_ABS}) with modes up to  $m=3$, although we resolve up to $N_m=20$, have an absolute value of the electron density $|n_e^m|$ that peaks at some intermediate radius $r$, which increases with $m$. In particular, the first mode peaks approximately at the edge of the injection area,
where the azimuthal homogeneity imposed by the uniform injection is no longer imposed.
The absolute value of the  electrostatic potential modes $|\phi^m|$ for each $m>0$ features a mild peak near the end of the injection zone.
Finally, the modal contributions $T_{\perp e}^m$
with $m>0$  are  negligible  in the injection region, meaning that the temperature remains essentially axisymmetric there, but increases in the peripheral region. 
Crucially, the $m>0$ modes of all these variables decay roughly as $1/m$ (including those not shown in the figure). 

However, since  $|\phi^m|$ decays with $1/m$, this means that the modal components of the azimuthal electric field $|E_\theta^m|$ remain almost constant with $m$, since
(cf. equation \eqref{eq:fielddef}):
\begin{align}
    |E_\theta^m| = \frac{m}{r} |\phi^m| .
\end{align}
This is evident in the plot of $| E_\theta^m |$: outside of the injection region and its neighborhood (where the modes do decay with $m$) the 
absolute value of the electric field is roughly the same for each $m$ mode.
This calls into question the feasibility of truncating the modal expansion at any finite $N_m$.

In practice, what determines the physical relevance of each mode is how it contributes to the wave-driven radial transport.  Said transport, as discussed more in detail in a later section \ref{sec:anomalous_tpt}, is given by the convolution in $\theta$ of the density and azimuthal electric field, so that each mode contribution scales as the product $n^m E^m_\theta$.
This does decay with $m$ thanks to the behavior of $n_e^m$. Hence, even if $|E_\theta^m|$ does not vanish with increasing $m$, it should still be possible to obtain accurate plasma profiles after truncation (provided a sufficiently high value of $N_m$).
Indeed, quickly-decaying transport contributions are essential to justify  the truncation of the modal expansion at a moderate $N_m$ for computational efficiency; 
this is further discussed in section \ref{sec:anomalous_tpt}; the convergence of the plasma profiles with varying $N_m$ is demonstrated in section \ref{sec:convergence}.
 
To complement our presentation of the simulation results, we briefly discuss the azimuthal electron velocity and the perpendicular electron pressure.
The axisymmetric $u^0_{\theta e}$ profile is  diamagnetic with respect to the applied field, and indeed, originates from the diamagnetic flux
$g_{\theta e,d} = - (\pd p_{\perp e} / \pd r)/(eB)$ and the $\bm E\times \bm B$ flux $g_{\theta e,E\times B} = - n_eE_r/(eB)$, except in the Larmor layer near the edge of the domain where it encounters additional (gyroviscous) contributions.
The $\bm E \times \bm B$ velocity is known to be one of the main triggers of drift-gradient instabilities in inhomogeneous plasmas \cite{smol17,ripo25a,ramo26a}, in particular of the collisionless Simon-Hoh instabilty \cite{saka93}. 
The resulting radial force $en_e^0u^0_{\theta e}B$ confines the electrons radially, and consequent with its origin, largely compensates the push of the radial pressure gradient $\pd p_{\perp e} / \pd r$ and the electric field $E_r^0$.
The modes $m>0$ of $u_{\theta e}$ roughly decay with $m$ in the first part of the domain, but, just like $E_\theta^m$, they become comparable in the second part. Again, the Larmor layer presents a different behavior than the bulk.

The modes of the perpendicular electron pressure $p_{\perp e}^m$ can be regarded as the convolution of the modes of the density $n_e^m$ and temperature $T_{\perp e}^m$
The dominance $T_{\perp e}^0 \gg |T_{\perp e}^m|$ (for $m>0$) in the first part of the domain means that the perpendicular electron pressure can be roughly approximated simply as $p_{\perp e}^m \simeq T_{\perp e}^0 n_e^m$. The linear reconstruction of $p_{\perp e}^m$ has been plotted as  dashed lines over the actual pressure profiles in figure \ref{fig:rfrnc_1D_ABS} to assess the validity of the approximation. This supports the isothermal closure frequently used in fluid linear instability literature \cite{smol17,ramo21,ripo25a}. However, the figure also shows that the approximation becomes inaccurate (especially for mode $m=1$) at larger radii.


The heavier, essentially-unmagnetized ions, whose profiles have not been plotted, present a density profile identical to that the electrons, as expected by the quasineutrality of the discharge. They are accelerated radially in the injection region, following from continuity $r^{-1}\pd (rn_i u_{ri}^0)/\pd r = \dot n_i$, and decelerated in the remaining of the domain due to the inward-pointing radial electric field $E_r^0= - \pd \phi^0 / \pd r$.

\subsection{Instantaneous plasma profiles}\label{sec:instantaneous}

The long-term averages above, while relevant to understand the mean plasma flow, conceal the low-frequency oscillations present in the modes. Moreover, the long averaging process prevents us from inspecting the time-varying phase of the modes.
Our \PIF\ approach is better suited to compute the short-term average  of the complex amplitudes of the azimuthal modes, which provides the near-instantaneous profiles of the plasma, while enabling us to perform enough statistics to characterize actual PIC noise. 
We discuss these next.

The short-term averages for the $m>0$ modal coefficients are computed in complex space using $N=2000$ consecutive timesteps centered about a specified time instant.  This corresponds to a physical time of $40$ ns (cf. with the 250 $\mu$s of the long-term averages). %
We define these short-term averages for the complex amplitude as:
\begin{align}
    \left| \langle n_e^m \rangle_s \right| = \left|\frac{1}{N}\sum_n^N n_e^m\right|,
    \qquad
    \sigma_s(n_e^m) =\sqrt { \frac{1}{N-1}\sum_n^N |\langle n_e^m \rangle_s - n_e^m|^2},
\end{align}
where we use the subindex $s$ when defining the short-term  averaging operator $\braket{\cdot}_s$ to differentiate it from its long-term  counterpart.
In addition, to correctly analyze the phase angle we employ circular statistics as in \cite{mard00,madd25a},
\begin{align}
    \langle\angle n_e^m\rangle_s = \angle \sum_n^N \exp \square{i \angle n_e^m }, \qquad
    \sigma_s(\angle   n_e^{m}) = \frac{1}{N-1} \sum_n^N \square{\pi - \Big|\pi - | \angle n_e^m - \langle \angle n_e^m \rangle_s|\Big|}.
    \label{eq:circular}
\end{align} 
%
Observe that, for cases with small deviation in the magnitudes of $n_e^m$, $\langle\angle n_e^m\rangle_s  \simeq \angle \langle n_e^m\rangle_s$ in practice. However, $\sigma_s(n_e^m)$ does not offer information on the standard deviation of the phase angle. For this we need the quantity defined in equation \eqref{eq:circular}, which is the  circular deviation of the phase of  $n_e^m$ \cite{mard00} and is the analogous to the standard deviation in linear statistics.

\begin{figure}
    \centering \includegraphics[width=0.8\linewidth]{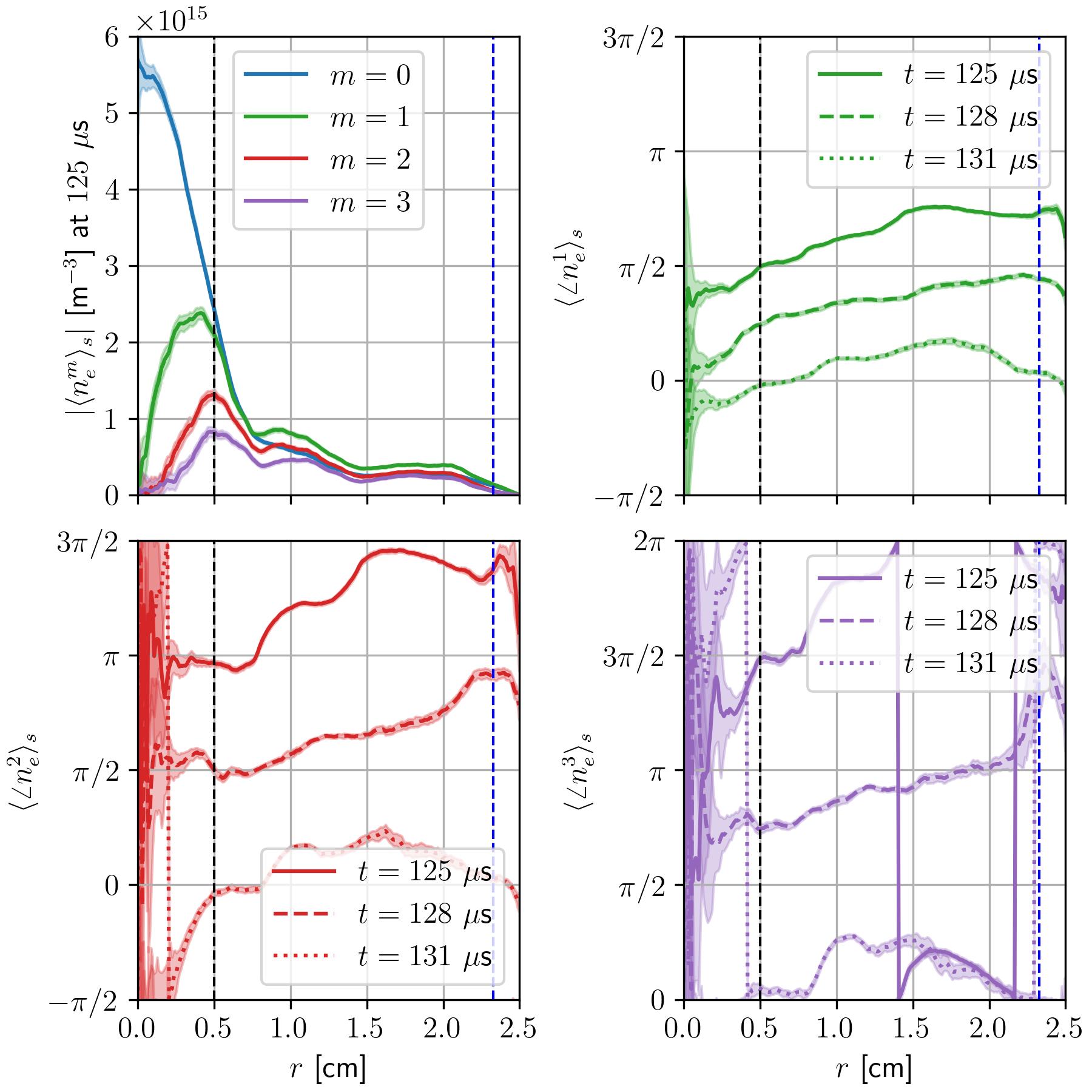}
    \caption{Short-time  mean and  deviation for the amplitude and phase of $n_e^m$ for $m=0,1,2,3$ at different time instants over $2000$ consecutive samples, for the reference simulation R. The plot in the upper left presents the amplitude of the average of ${n_e^m}$ at $t=125 \ \mu$s. The other three plots present the circular mean and deviation of the phase of the first three modes at $t=125$, $128$ and $131 \ \mu$s.
    For the meaning of the vertical dashed lines see the caption of figure \ref{fig:rfrnc_1D_ABS}.}
    \label{fig:instantaneous}
\end{figure}

Figure \ref{fig:instantaneous} presents the counterpart of figure \ref{fig:rfrnc_1D_ABS} but for the short-term averages, evaluated at $t=125$, $128$, and $131 \ \mu$s, which correspond with the time instants shown in the snapshots of figure \ref{fig:rfrnc_2D}.
These plots show that a short-time average with a small $\Delta t$ has a much narrower standard deviation (which in this case would come solely from statistical noise, and not from physical oscillations over long periods of time) than those displayed in figure \ref{fig:rfrnc_1D_ABS}.
The phase angle of the different modes describes the rotation of the spoke.
The phase of each mode (specially $m=1$) increases slightly with $r$, which is consistent with the curved spoke structure observed in figure \ref{fig:rfrnc_2D}. 
The phase rate scales roughly with $m$, which gives an angular velocity of about $m \pi / 2$ over $6 \ \mu$s for the $m$-th mode, corresponding to an approximate spoke frequency of $\sim m\times 48$ kHz, compatible with the results later obtained in the frequency analysis of section \ref{sec:freqdomain}.

Incidentally, while for the long term averages of section \ref{sec:long-term}, at the axis $\phi^m\to 0$ (for $m>0$),
the nonsymmetric modes of $n_e$ and $\phi$ do not exactly vanish there; this is an artifact of numerical noise and the fact that we are averaging $|n_e^m|$, $|\phi^m|$, and not the full complex magnitudes (whose average does go to zero), as corroborated in this section.

\subsection{Frequency domain analysis}\label{sec:freqdomain}

The azimuthal mode decomposition in the \PIF\ scheme has the intrinsic advantage that it already outputs each modal component
of the rotating structures, which are frequently required for further analysis. In this section we briefly illustrate
how we can gain  further insight into the dynamics of the spoke by transforming the time series of each mode profile into the frequency domain.

For each $m$, the modal coefficient in equation \eqref{eq:fourierexpansion} can be expressed as
\begin{align}
    n_e^m (r,t) = \sum_{n=-N_T/2}^{N_T/2 - 1} n_e^{m,n} (r) \exp\round{-i \w^n t}
\end{align}
with $\w^n = 2 \pi n/ (N_T \Delta t_{p})$ the discrete $n$-th frequency, and $N_T$ is the even number of print steps,
spaced by $\Delta t_{pr}$.
Now we must retain both positive and negative frequencies in $n_e^{m,n}$, as $n_e^m$
is complex (with the exception for $m=0$, which has a symmetric time spectrum).
The power spectral density (PSD) of a given plasma quantity, say $n_e$, is defined as the squared norm of $n_e^{m,n}$, and it is computed by averaging it over different windows and indicated as PSD(${n_e^m}$).
In this work we select $10$ time windows between $t=156.25$ and $500$ $\mu$s with an overlap of $50 \%$ each, downsampling the data to $N_T=1250$ print steps with $\Delta t_{pr}=5 \times 10^{-2} \ \mu$s, which allows resolving the relevant frequencies.
The frequency bin size under these parameters is $16$ kHz.

\begin{figure}
    \centering
    \includegraphics[width=.45\linewidth]{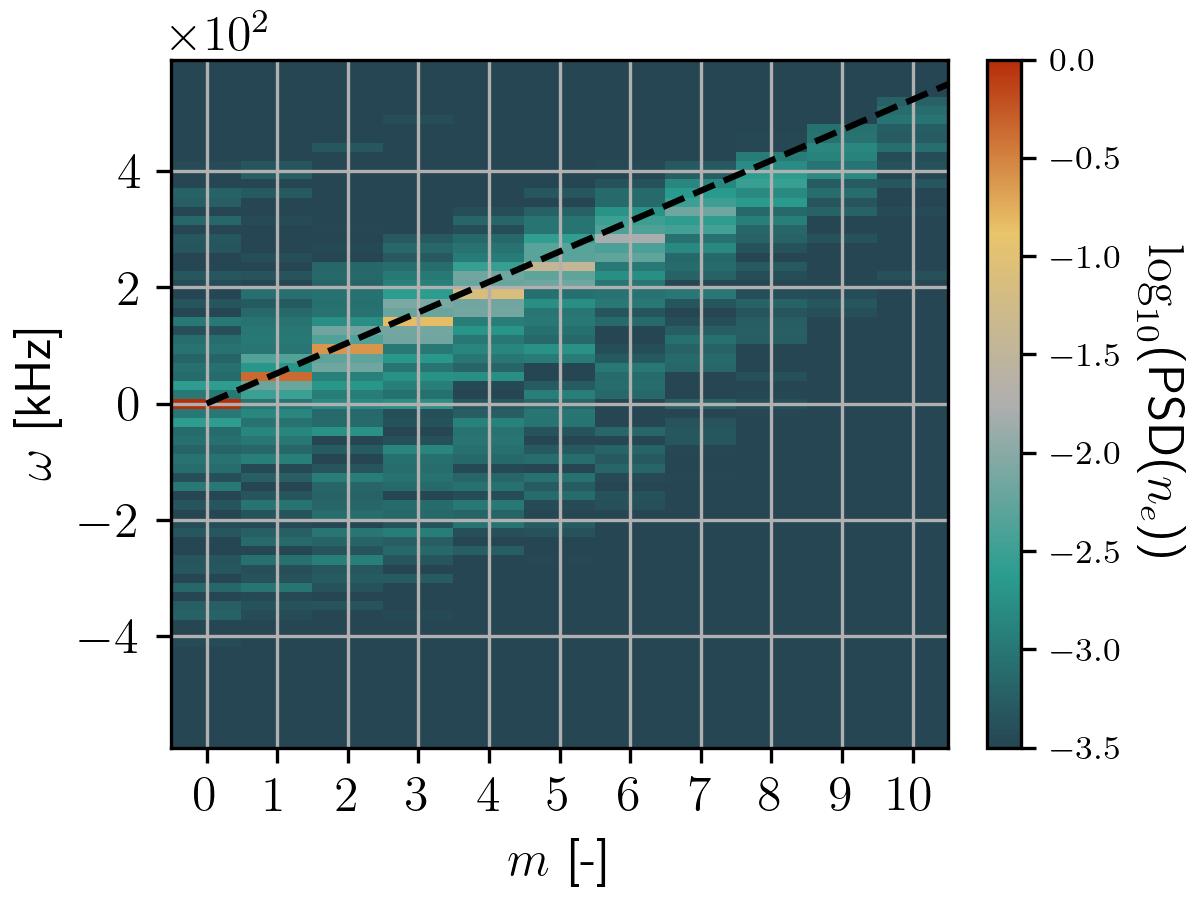}
    \caption{
    PSD of $n_e^m$ for the first azimuthal modes of the reference simulation R at $r=1.25$ cm. 
    Values have been normalized with respect to the peak PSD value overall. The dashed black line indicates the averaged sonic velocity, $\braket{c_s} $.}
    \label{fig:rfrnc_KW}
\end{figure}

Figure \ref{fig:rfrnc_KW} presents the normalized PSD of $n_e^m$ at $r=1.25$ cm for $m$ up to 10. The normalization has been carried out using as a reference the maximum overall PSD value, which takes place for the $m=1$ mode at approximately $
48$ kHz. 
It is evident that the PSD peaks on a diagonal line with a single phase velocity, $v_{\theta} = r\w / m$, consistent with a rotating spoke. Interestingly, the velocity $v_{\theta}$ matches the  ion sound velocity computed at $r=R$, $\langle c_s \rangle =(\langle T^0_{\perp e}\rangle/m_i)^{1/2}$; this could indicate that the observed structure might originate due to an instability taking place at the domain edge, where the $E_r \times B_z$ drift is larger and the collisionless Simon-Hoh instability growth rate $\gamma_{SHI}$ is larger, as $\gamma_{SHI} \propto u_{E\times B}$ \cite{smol17,ripo25a,ramo26a}. In \cite{powi26} the spoke fluctuation is justified by computing the local value of the real frequency of the Simon-Hoh unstable mode, approximately $\simeq 53$ kHz. While this value aligns well with our observed phenomenon, it could be argued that a local and linear estimate does not constitute a completely valid argument for a global, saturated wave such as the one simulated.
The PSD of $\phi$ (not shown) displays the same structure, revealing that the electrostatic potential follows similar as the density spoke.

Incidentally, the spectrum in figure \ref{fig:rfrnc_KW} also reveals another interesting feature of the simulation. While the spectrum of each mode $n_e^m$ displays a clear peak, there is non-negligible spectral content at other frequencies too. This is an indication that the modal amplitudes cannot be simply described as a constant absolute value plus a phase angle that varies linearly in time. Indeed, this is consistent with the behavior of the instantaneous profiles discussed in section \ref{sec:instantaneous},
where the absolute value of the modal amplitude is seen to vary slowly in time; moreover, the phase does not vary exactly linearly.
These oscillations occur over a slow time scale (corresponding to the frequencies in the spectrum), and clearly differ from statistical noise oscillations.

\subsection{Cross-field electron transport}\label{sec:anomalous_tpt}

As indicated above, a major concern in all simulations of low temperature plasma sources is the (collisionless) cross-field transport rate of particles to the lateral walls. 
Correctly simulating cross-field transport is essential for an accurate simulation, as the plasma density, electric potential, and other variables depend strongly on it. Moreover, plasma losses to the wall are a major component of source inefficiencies, whose characterization is typically of great interest.
For this reason, in this section we illustrate how the \PIF\ scheme naturally isolates the contribution to cross-field transport of each mode, making it specially suited for this type of studies.

While the heavier ions are guided essentially by the electric field $\bm E$,
the lighter, magnetized electrons respond both to $\bm E$ and the applied magnetic field $\bm B$, and to leading order approximation in their Larmor radius expansion, individual electrons merely move in the (instantaneous) $\bm E\times \bm B$ direction. 
The azimuthal asymmetries in the discharge, which enable the development of $E_\theta$, are 
what allows a radial electron current to flow from the injection region to the boundary; this essentially non-symmetric phenomenon is generally referred to as anomalous transport, and it can greatly surpass classical collisional transport in many devices \cite{kaga20a}.

\begin{figure}
    \centering
    \includegraphics[width=0.8\linewidth]{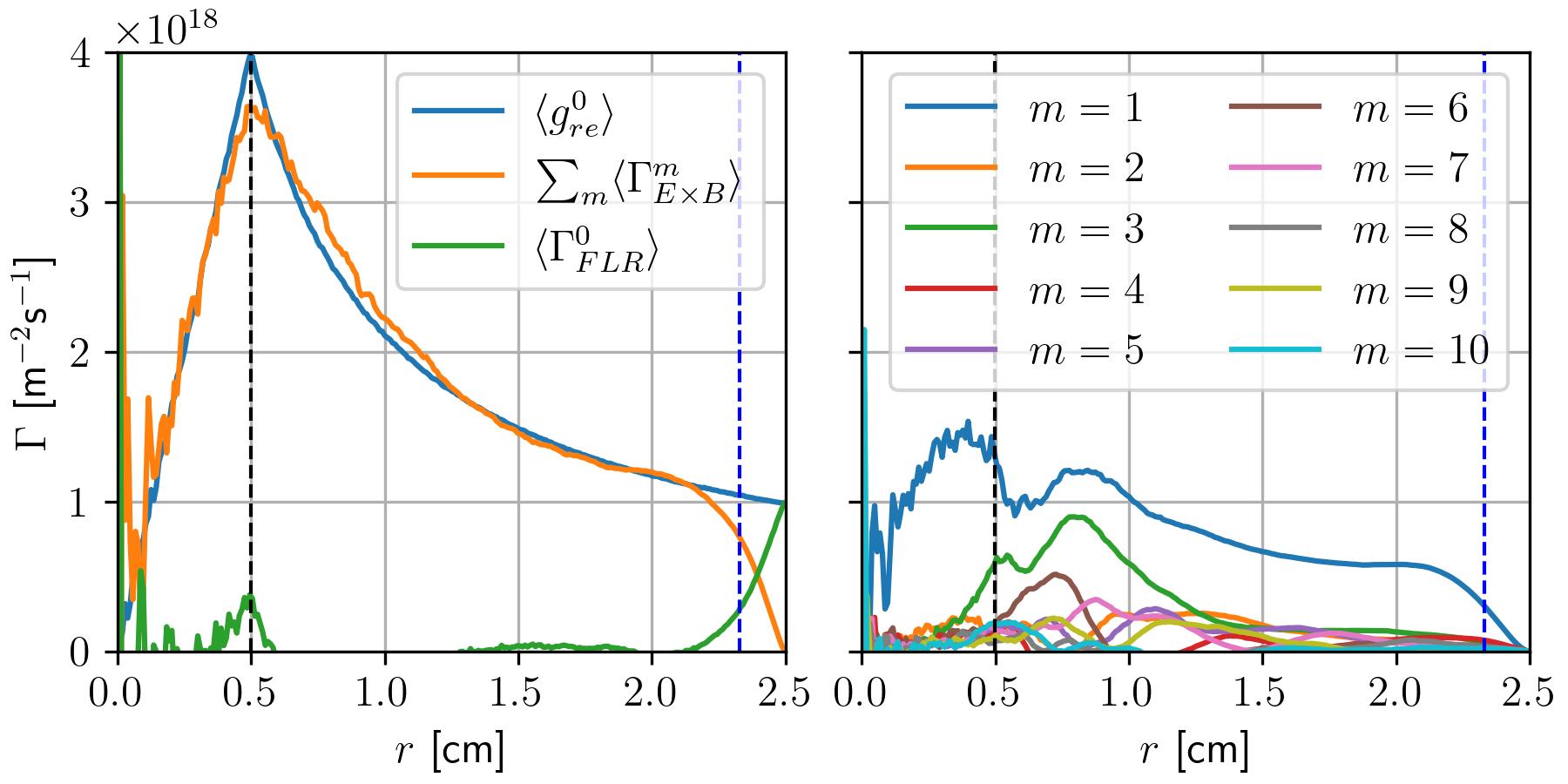}
    \caption{Left: long-time-averaged electron flux in the reference simulation R. Right: components from $m=1$ to $m=10$ of the azimuthal anomalous transport terms, $\Gamma^m_{E\times B}$.
    For the meaning of the vertical dashed lines see the caption of figure \ref{fig:rfrnc_1D_ABS}.}
    \label{fig:currents}
\end{figure}

In our collisionless plasma, the radial electron transport contributed by each azimuthal mode is best understood by inspecting the complete azimuthal electron momentum equation,
\begin{align}
g_{re} eB = en_eE_\theta + m_e\frac{\pd g_{\theta e}}{\pd t} + \round{\nabla \cdot \mathcal M_e}_\theta.
\label{eq:electron_mom}
\end{align}
where $\bm g_{e} = n_e\bm u_{e}$ is the electron flux density, 
and $\mathcal M_e$ is the complete electron momentum tensor (including inertia, pressure, and gyroviscosity), as in section \ref{sec:long-term}.
Averaging in $\theta$ yields the equation for $g_{re}^0$, the net axisymmetric radial electron transport. 
The only component of $\mathcal M_e$ that survives averaging is the part of $\mathcal M_{r\theta e}^0$ 
associated to inertia and gyroviscosity, which are both first order in the electron Larmor radius expansion.
Next, long-term averaging in time ($\langle\cdot\rangle$) eliminates the
temporal derivative in \eqref{eq:electron_mom}.
Finally, writing the $m=0$ mode of the $n_eE_\theta$   term as the convolution of higher order modes of $n_e$ and $E_\theta^m=-im \phi^m/r$, after some algebra equation \eqref{eq:electron_mom} becomes
\begin{align} 
\langle g_{re}^0 \rangle  &= 
\sum_{m=1}^{N_m} \Gamma^{m}_{E\times B}
+ 
\Gamma_{FLRE}^0, 
\label{eq:electron_mom_avg}
\end{align}
with
\begin{align}
\Gamma_{E\times B}^{m} &= 
\frac{m}{2rB}
\left\langle 
|n_e^m||\phi^m|\sin \alpha^m
\right\rangle 
= \frac{m}{2rB}\Im\left[\mathrm {CSD}(n^m_e , \phi^m)\right],
\\
\Gamma_{FLR}^0 &=
\frac{1}{eB}\left
\langle
\frac{\pd \mathcal{M}_{r\theta e}^0}{\pd r} + \frac{2\mathcal{M}_{r\theta e}^0}{r} 
\right\rangle,
\end{align}
where we have introduced $\alpha^m=\angle \phi^m - \angle n_e^m$, 
the relative instantaneous phase angle between  $n_e^m$ and $\phi^m$. The terms  $\Gamma_{E\times B}^m$ are the contribution to anomalous transport of each $m$ mode, and $\Gamma_{FLR}^0$ captures the contribution of finite Larmor radius effects.
Note that the angle bracket in the definition of $\Gamma_m$ coincides with the imaginary part of the cross spectral density (CSD) between $n_e^m$ and $\phi^m$.  
Figure \ref{fig:currents} depicts $\langle g_{re}^0 \rangle$ and its two mechanisms ($\bm E \times \bm B$ transport and inertia/gyroviscosity), plus the profiles of the modal decomposition $\Gamma^m_{E\times B}$.
We note that the shape of $\langle g_{re}^0 \rangle$  is known a priori: since the injection profile is known, and $\langle g_{re}^0 \rangle$ must balance this input exactly at steady state, $\langle g_{re}^0 \rangle$ is indeed prescribed; observe that, beyond $r_{inj}$, the geometrical expansion of the radial electron flux $\propto 1/r$ becomes evident.

What the \textit{effectiveness} of the transport mechanisms determines is the level of the electron density $n_e$ and the potential $\phi$, as well as their oscillations, at steady state: a highly-efficient transport mechanism will result in an overall lower electron density profile, whereas a large one results if transport is strongly impeded.

Our results demonstrate that the totality of the electron flux is enabled by the wave-driven transport mechanism everywhere,
with the exception of a Larmor layer in  the neighborhood of the wall, where the electron distribution becomes highly asymmetric in the gyrophase as electrons get collected by the wall, and finite Larmor radius phenomena have significant net effect.
The breakdown of anomalous transport into its modal contributions $\Gamma^m_{E\times B}$ indicates that $m=1$ has the leading contribution, while all other modes have a small yet nonzero additional one. 
In particular, $m=3$ yields an important effect on transport right outside the injection area, with the sum of $m=1$ and $m=3$ modes contributing  $\sim 50 \%$ of the total maximum radial transport reached at $r=r_{inj}$.

\begin{figure}
    \centering
    \includegraphics[width=0.45\linewidth]{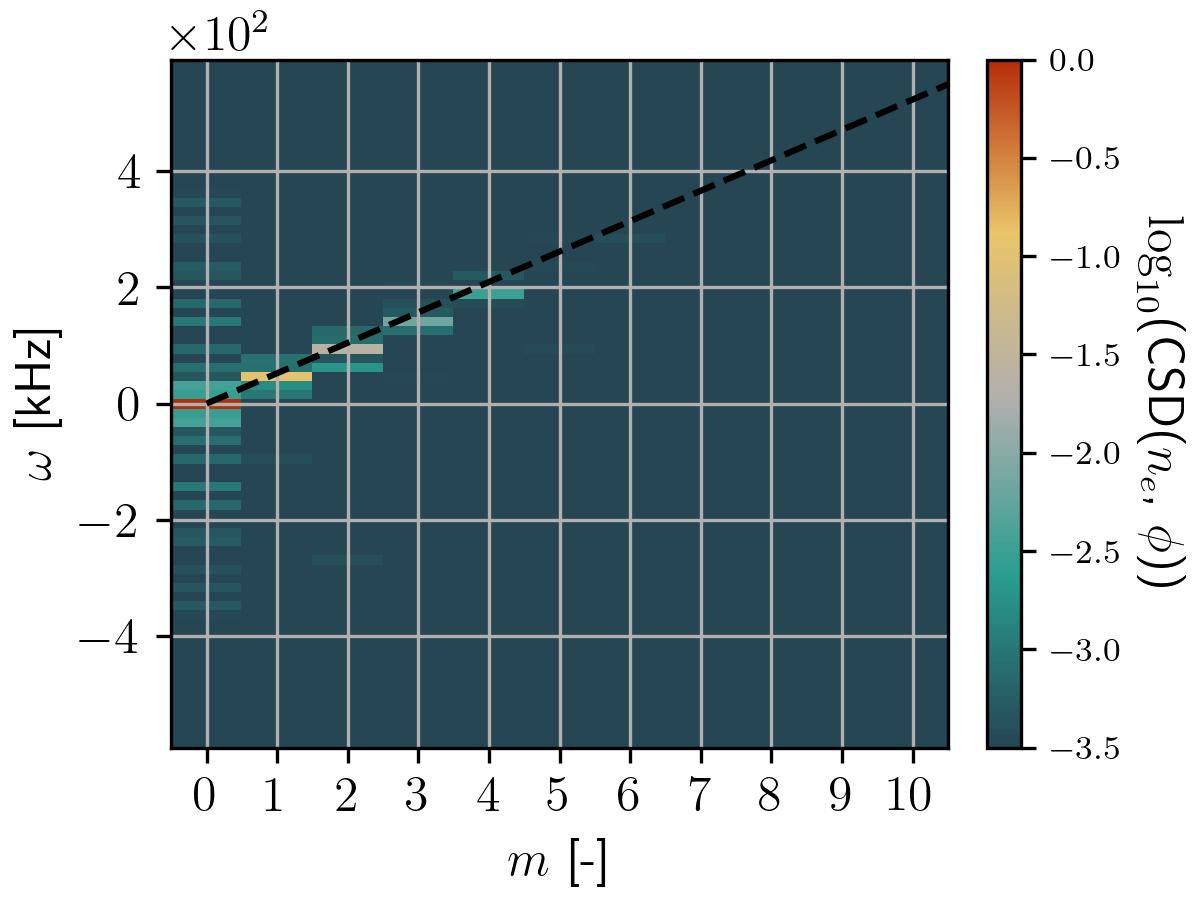}
    \includegraphics[width=0.45\linewidth]{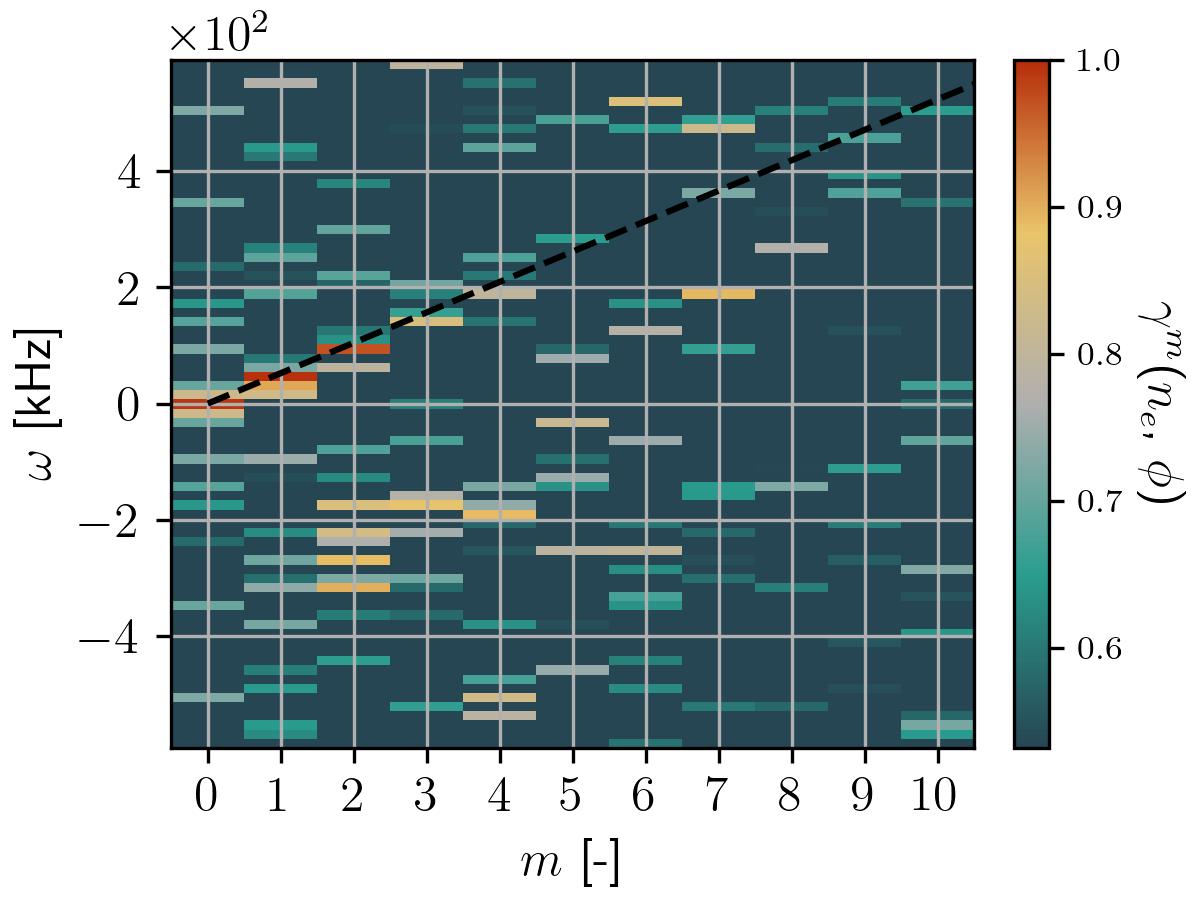}
    \caption{Cross spectral density CSD and coherence $\gamma^m$ between $n_e$ and $\phi$ in the $(m,\w)$ space measured at $r=1.25$ cm for the reference simulation R. For the meaning of the dashed black line see the caption of figure \ref{fig:rfrnc_KW}.}
    \label{fig:CPSD}
\end{figure}

To conclude this section, we additionally inspect the CSD between $n_e^m$ and $\phi^m$, 
which measures the consistency of the  phase coupling between
these two variables, essential to yield a nonzero average $\Gamma^m_{E\times B}$ over time.
In addition, we present  its normalized version, the coherence,
\begin{align}
\gamma^m = \frac{|\mathrm{CSD}(n_e^m,\phi^m)|}{\sqrt{\mathrm{PSD}({n_e^m})\mathrm{PSD}({\phi^m})}}.
\end{align}
For a given number of time windows $N_w$, the coherence level at which we reach minimum statistical significance is given by $(1-0.05^{1/(N_w-1)})^{1/2}$ as reported in \cite{madd25a}. In our case with $N_w=10$ windows, the threshold corresponds to $0.532$.
This is shown in figure \ref{fig:CPSD}, which indicates that the strongest coupling of $n_e$ and $\phi$ occurs for modes $m=1$ to $5$ at the frequencies of the spoke; nevertheless, the coherence reveals that other statistically-relevant couplings exist outside of this diagonal line.

\section{Numerical convergence analysis}
\label{sec:convergence}

In this section we explore the effect of varying the numerical parameters on the 
plasma profiles. This allows us to demonstrate that our reference simulation can indeed be considered `converged', and to understand the computational cost of the scheme. Relatedly, this allows us to determine whether some of the parameters can be relaxed with a small effect on fidelity.

\begin{figure}
    \centering
    \includegraphics[width=.45\linewidth]{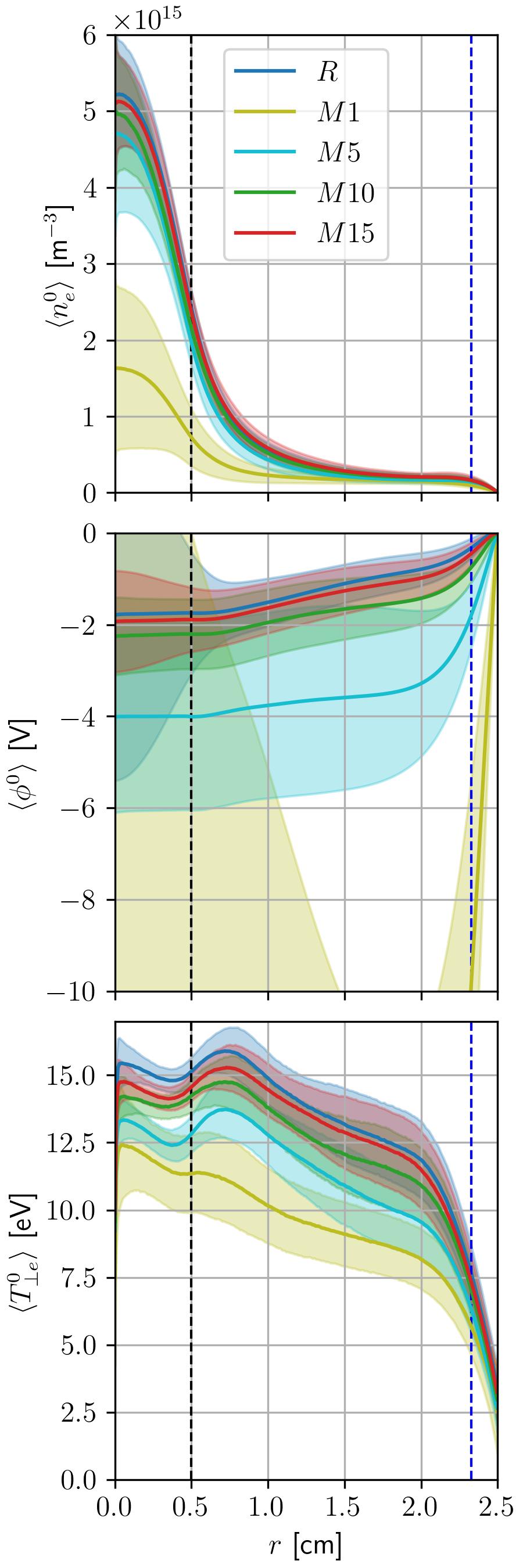}
    \includegraphics[width=.45\linewidth]{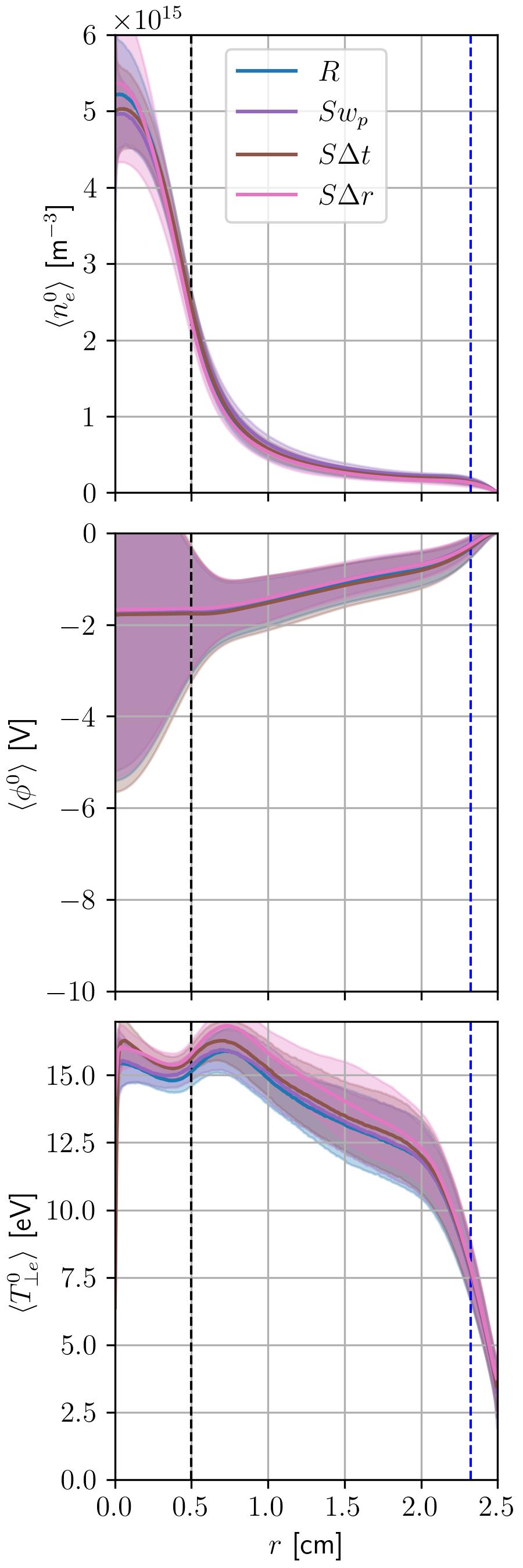}
    \caption{Comparison between long-time averaged radial profiles and respective standard deviations between the reference simulation case R and the additional simulations of the convergence study. Vertical dashed lines are as in figure \ref{fig:rfrnc_1D_ABS}.}
    \label{fig:convergence}
\end{figure}

A total of $7$ additional simulation cases are listed in Table \ref{tab:cases}, and their $m=0$ profiles are shown in figure \ref{fig:convergence}.
Simulations M1, M5, M10 and M15 explore the effect of varying $N_m$.
Simulation S$w_p$ doubles the reference particle weight, which roughly diminishes the number of particles per cell $N_p$ by half.
Simulation S$\Delta t$ doubles the reference timestep.
Lastly, simulation S$\Delta r$ increases the reference cell size.
Note that, in keeping $w_p$ constant in this last case, the number of particles per cell is also increased, while the total number of particles in the domain remains essentially the same.

\begin{table}
    \centering
    \begin{tabular}{|c|c|c|c|c||c|c|}
    \hline
    Case & $N_m$ & $w_p$ & $\Delta t$ & $\Delta r$ & Wall time & Total particles \\
     & & [$10^6$] & [$10^{-2}$ ns] & [$10^{-2}$ cm] & [h] & [$10^6$] \\
     \hline
     \hline
    R     & $20$ & $2$ & $2$ & $1$ & $110$ & $1.8$ \\
    \hline
    \hline
    M1     & \textbf{1} & $2$ & $2$ & $1$ & $81$ & $1.2$ \\
    \hline
    M5     & \textbf{5} & $2$ & $2$ & $1$ & $95$ & $1.6 $ \\
    \hline
    M10     & \textbf{10} & $2 $ & $2$ & $1 $ & $103$ & $1.7 $ \\
    \hline
    M15     & \textbf{15} & $2$ & $2$ & $1$ & $108$ & $1.8$ \\
    \hline
    S$w_p$     & $20$ & $\mathbf{4}$ & $2$ & $1$ & $59$ & $0.8$ \\
    \hline
    S$\Delta t$     & $20$ & $2$ & $\mathbf{4}$ & $1$ & $65$ & $1.8$ \\
    \hline
    S$\Delta r$   & $20$  & $2$ & ${2}$ & $\mathbf{2}$ & $114$ & $1.9$ \\
    \hline
    \end{tabular}
    \caption{Additional simulation cases and their respective numerical parameters for the convergence analysis. All simulations have been run on a single Intel Xeon Gold Y6542 CPU (using only 16 physical cores) and RAM of $256$ GiB.
    }
    \label{tab:cases}
\end{table}

As discussed in section \ref{sec:anomalous_tpt},
the maximum number of simulated modes $N_m$  influences radial transport and affects the average plasma profiles.
%
Given the fact that the first modes are the dominant contributors to radial transport, it is tempting to explore how much information is lost if we perform a more economic simulation with a lower value of $N_m$, or perhaps even $N_m=1$. As $\langle g_{re}^0 \rangle$ is fixed by injection, the magnitude of the modal oscillations and their phase coupling in this new case must increase, such that the $E\times B$ induced transport matches the required profile in steady state. 


The left column of figure \ref{fig:convergence} displays $\langle n_e^0 \rangle$, $\langle \phi^0 \rangle$ and $\braket{T_{\perp e}^0}$ for three additional simulations, identical in all aspects to the reference R simulation, but with $N_m=1,5,10,15$.
As the number of simulated modes $N_m$ decreases, the axisymmetric density profile $\langle n_e^0 \rangle$ at steady state decreases, the potential well in $\phi^0$ becomes deeper, and the electron temperature $T_{\perp e}^0$ decreases. Meanwhile, the standard deviation of all three variables increases, specially for $\phi^0$.
Pushing this exercise further to the case of purely axisymmetric simulation (i.e., with $N_m=0$) leads to a crash in the simulation as no steady state is reached; the reason stems from the fact that 
electrons cannot leave the domain if there are only axisymmetric electric fields, even if time-varying, unless they reach the keV-level energies that are required to have a Larmor radius comparable to the domain size. Before that can occur, the negative space charge grows and the resolution constraints of section \ref{sec:constraints} are eventually violated. I

Simulations M1, M5, M$10$, M$15$, and R
demonstrate that convergence has already been reached in practice with 10 modes. The addition of ulterior modes adds little further information to the physics of the problem.
However, while M$5$ shows similarities in the density and temperature profiles, M$1$ differs importantly from the other cases for the reasons discussed in the previous section \ref{sec:anomalous_tpt}. Both these last two cases show an increasingly large potential well in $\phi^0$ with large standard deviations, with case M$1$ reaching a minimum of $\braket{\phi^0}\simeq -30$ V.
Intuitively, in these simulations the plasma is deprived of enough modes to fully develop the necessary radial $E\times B$ electron transport; as ions escape the domain the potential well deepens and its oscillations in time (characterized by the standard deviation) grow. The large, time-varying potential eventually enables the electrons to outflow, but this differs substantially from the physical situation observed in well resolved simulations and in the benchmark of \cite{powi26}. 
This illustrates the need to explore the numerical convergence of the simulation with parameter $N_m$ in each problem, to identify an appropriate cutoff value as a function of the desired level of accuracy.


Lastly, the right column of figure \ref{fig:convergence} demonstrates that our chosen numerical parameters for simulation R yield convergent results under variations of $\Delta r$, $\Delta t$, and $w_p$, and that, very likely, these parameters can be relaxed without a major sacrifice in accuracy,
while satisfying the constraints established in section \ref{sec:constraints}.


Regarding the computational cost,
Table \ref{tab:cases} presents the performance in wall time of the various \PIF\ simulations, where the near-linear scaling with $1/w_p$ and $1/\Delta t$ is evidenced.
%
There is a much weaker dependency on the number of modes $N_m$, with cases R, M1--M15 all sharing a similar run time. The reason for this is the efficient computation of the trigonometric functions explained in section \ref{sec:model}, wherein most of the cost is paid in the computation for mode $m=1$ already, and additional modes incur in a small additional cost only.
Likewise, the dependency on $\Delta r$ is small, as the Poisson solver in 1D has a negligible cost. 
  
While we note that no special provisions were taken to optimize the present implementation, these data show a promising lower computational runtime when compared to \PIC \ simulations of the same physical problem.
Indeed, in the case of the simulation with 2D \texttt{PICASO}
in \cite{powi26}, simulation wall time was $181$ h (7 d 13 h) using  2 Intel Xeon Silver 4316 CPU using all 40 physical cores, and  other codes therein completed the simulation in as much as 55 days with other hardware.

Crucially, those simulations utilized  $\Delta t=4\times 10^{-11}$ s (double our reference timestep) and $\Delta x = 2 \times 10^{-2}$ cm (double our reference grid spacing).
Compared with the 2D \texttt{PICASO} simulation and accounting for these differences (and the fact that that simulation ran in 40 cores rather than 16), 
the present \PIF\ simulation results in a reduction factor of $\sim 8$--$10$ for the same time resolution. This is consistent with the lower total number of macroparticles in our simulation, which is also around a factor of $10$.
%

\section{Conclusions}
\label{sec:conclu}




We have presented a Fourier-decomposed Particle-in-cell algorithm to resolve low-temperature cylindrical plasma discharges, and have illustrated its performance in  a
two-dimensional collisionless Penning discharge case.
Our treatment expands all spatial functions (fields, densities, higher moments of the distributions) in Fourier modes in the azimuthal direction, while it retains a grid in the remaining directions, which facilitates the imposition of complex boundary conditions. Particle pushing and weighting remains full-dimensional.
The modal decomposition in the azimuthal direction naturally exploits the cylindrical geometry of the source, provides spectral accuracy in this direction, and brings in the possibility of truncating the computation at a chosen number of modes $N_m$. Furthermore, it directly outputs the quantities of interest in many analyses (namely, the modal amplitudes themselves).

The \PIF\ simulation of the Penning discharge displays the well-known density spoke, which rotates with the characteristic sonic velocity. 
In spite of the differences in domain shape, reasonable agreement is found between our model and the community benchmark results of \cite{powi26}. 
Modes $m>0$ decay as $1/m$ in a way that means their contribution to instability-driven cross-field transport decreases with $m$, an essential feature to justify the modal truncation.

We have analyzed the influence of the total number of azimuthal modes $N_m$ and particle count $N_p$ on the overall cross-field transport and the computational time.
The scheme has the potential to bring relevant computational savings with respect to traditional finite-differencing PIC.
The major cost saving comes from reducing 
the total number of particles in the simulation, as the two-dimensional grid is broken down into a multitude of one-dimensional grids, one per mode.
Furthermore, our computational cost analysis shows that, at least in 2D, the cost of adding additional modes is rather marginal thanks to the efficient storage and computation of the trigonometric functions involved. 

\section*{Acknowledgments}

This project has received funding from the European Research Council (ERC) under the European Union’s Horizon 2020 research and innovation programme (Starting Grant project ZARATHUSTRA, grant agreement No 950466).   
Additional support came from the R\&D project PID2023-150052OB-I00 (ADAPT) funded by MICIU/AEI/ 10.13039/ 501100011033 and by ERDF, EU.

The participation of E. Ahedo lies within the R\&D project PID2022-140035OB-I00 (HEEP) funded by MCIN/AEI/10.13039/501100011033 and by “ERDF A way of making Europe.”

The Department of Aerospace Engineering at Universidad Carlos III de Madrid has been accredited as a Unit of Excellence “María de Maeztu”, a distinction awarded by the Spanish State Research Agency for the period 01/01/2026-31/12/2031.


\appendix

\section{Potential estimation at the first node}

\label{app:noise}

As discussed in section \ref{sec:constraints}, we require an accurate estimate of the potential close to the axis to assess the magnitude of noise-driven particle acceleration. In particular, we are interested in finding how the potential modes $\phi^m$ scale with their azimuthal mode number $m$. For each Fourier mode, the potential can be expressed by means of the Green's function of the Laplacian as
\begin{align}
    \phi^m(r) = - \frac{1}{\varepsilon_0} \int_0^R G^m (r,r') \rho^m (r') r' dr' .
    \label{green:int}
\end{align}
For the boundary conditions reported in \ref{sec:Penning}, i.e. $\phi^m=0$ at $r=0,R$, the Green's function for $r<r'$ is
\begin{align}
    G^m(r,r'>r) = \frac{1}{2 m} \square{1 - \left( \frac{r'}{R} \right)^{2m}} \left( \frac{r}{r'} \right)^{m} \simeq  \frac{1}{2 m}  \left( \frac{r}{r'} \right)^{m},
    \label{green}
\end{align}
where the latter approximation is valid for $r' \ll R$.
Discretizing Eq. \eqref{green:int} on the radial grid for $r=\Delta r$ with $r'=j\Delta r$ gives 
\begin{align}
    \phi^m(\Delta r)
    \simeq
    - \frac{\Delta r^2}{2m\varepsilon_0}
    \square{\rho^m(\Delta r) + \frac{\rho^m(2\Delta r)}{2^{m-1}} + \frac{\rho^m(3\Delta r)}{3^{m-1}} + \cdots}    .
\end{align}
The contributions from $r'= j \Delta r$, $j\gg 1$, are increasingly smaller.
For large enough $m$, 
all contributions with $j>1$ can be neglected in first approximation, and the potential reduces to
\begin{align}
    \phi^m(\Delta r)
    \approx
     - \frac{\Delta r^2}{\varepsilon_0}     \frac{\rho^m(\Delta r)}{2m}.
    \label{eq:firstnodephi}
\end{align}
Substituting $\delta \rho^m$
in Eq. \eqref{eq:firstnodephi} then leads to the form for $\delta \phi^m$ presented in Eq. \eqref{eq:noise}.

\begingroup
\bibliographystyle{ieeetr} 
\bibliography{bibtex/ep2,bibtex/others}
\endgroup

\end{document}